\documentclass[aps,prd,reprint,nofootinbib,superscriptaddress,floatfix]{revtex4-2}

\usepackage{amsmath,amssymb,bm}
\usepackage{graphicx}
\usepackage{booktabs}
\usepackage{array}
\usepackage{xcolor}
\usepackage{hyperref}
\usepackage{siunitx}
\usepackage{microtype}
\usepackage{mathtools}
\usepackage{enumitem}
\usepackage{multirow}
\usepackage{url}

\hypersetup{
  colorlinks=true,
  linkcolor=blue,
  citecolor=blue,
  urlcolor=blue
}

\newcommand{\bt}{b_T}
\newcommand{\kt}{k_T}
\newcommand{\qT}{q_T}
\newcommand{\FNP}{F_{\mathrm{NP}}}
\newcommand{\NthreeLL}{\mathrm{N}^{3}\mathrm{LL}}
\newcommand{\GeV}{\mathrm{GeV}}

\newcommand{\as}{\alpha_s}
\newcommand{\mub}{\mu_b}
\newcommand{\zetaB}{\zeta_b}

\newcommand{\paperfigure}[2]{\includegraphics[width=#2\linewidth]{#1}}

\begin{document}

\title{Deep-neural-network extraction of unpolarized transverse-momentum-dependent parton distributions in $b_T$ space from Drell--Yan data}

\author{I.\ P.\ Fernando}
\email{ishara@virginia.edu}
\affiliation{Department of Physics, University of Virginia, Charlottesville, Virginia 22904, USA}

\author{D.\ Keller}
\email{dustin@virginia.edu}
\affiliation{Department of Physics, University of Virginia, Charlottesville, Virginia 22904, USA}

\date{August 25, 2026}

\begin{abstract}

We present a physics-informed deep-neural-network extraction of unpolarized transverse-momentum-dependent parton distribution functions (TMDPDFs) in impact-parameter space from Drell--Yan data.  The perturbative contribution is computed with a resummed $W$ term using $\mathrm{N}^{3}\mathrm{LL}$ evolution, strict-NLO hard and operator-product-expansion matching, and smooth profile scales at small and large $b_T$.  A compact feature-wise linear modulation network learns only a shared nonperturbative factor  $F_{NP}(x,b_T)$; the collinear PDFs, hard factor, evolution kernel, matching coefficients, and Fourier--Bessel transform remain fixed.  
The primary result is a smooth light-flavor $b_T$-space TMD ensemble and its cross-section-level validation.  The reported $k_T$ distributions are regularized finite-$b_T$ Hankel transforms, not independent momentum-space fits.  As a separate robustness test, a smooth finite-$Y$ transition is applied to 24 additional Tevatron points extending to $q_T/Q\simeq0.30$.  The nominal 329-point fit is unchanged, and the results remain stable when $F_{\rm NP}$ is held fixed while the transition profile is varied.  An independent 122-bin Tevatron $\mathrm{N}^{3}\mathrm{LL}+\mathrm{NNLO}$ $W+Y$ grid provides a direct perturbative benchmark. A separate $W+Y$ candidate using the specified non-LHCb finite-$Y$ inputs is retained as an identifiability study.
\end{abstract}

\maketitle

\section{Introduction}
\label{sec:introduction}

Transverse-momentum-dependent parton distribution functions (TMDPDFs) provide a three-dimensional description of hadron structure by resolving partonic longitudinal momentum and transverse motion.  They enter the factorization of low-transverse-momentum Drell--Yan (DY), semi-inclusive deep-inelastic scattering (SIDIS), and electroweak-boson production, and form the unpolarized baseline against which spin-dependent TMDs are normalized and interpreted \cite{CollinsSoper1981,CollinsSoper1982,CSS1985,CollinsBook}.

Impact-parameter space is the standard representation for perturbative TMD evolution.  The Fourier--Bessel transform converts transverse convolutions into products, while logarithms generated by the hierarchy between the hard scale $Q$ and the impact-parameter scale $\mu_b\sim 1/b_T$ are resummed through Collins--Soper and renormalization-group evolution.  Modern analyses have reached NNLL and $\NthreeLL$ logarithmic accuracy and have demonstrated that fixed-target and collider DY data constrain both the perturbative evolution and the nonperturbative large-$\bt$ sector \cite{Bacchetta2020,Bacchetta2022}.  At the same time, the functional form used for the nonperturbative contribution remains an important source of model dependence.

The current phenomenological landscape is characterized by several
complementary global-analysis programs.  Joint DY--SIDIS analyses now determine
unpolarized TMDPDFs, TMD fragmentation functions, and the Collins--Soper kernel
within a common fit, with subsequent studies resolving quark-flavor and
hadron-species dependence and propagating uncertainties from the corresponding
collinear distributions
\cite{ScimemiVladimirov2020,Bacchetta2022,BacchettaFlavor2024}.  The
perturbative treatment has also continued to advance, with a recent analysis
reaching $\mathrm{N}^{4}\mathrm{LL}$ accuracy while incorporating large-$x$
resummation and studying transverse-momentum moments of the extracted
distributions \cite{Moos2025}.  Simultaneous determinations of TMD and
collinear PDFs, together with joint experimental--lattice extractions of the
Collins--Soper kernel, are beginning to address correlations among the
transverse, collinear, and evolution sectors more directly
\cite{BarryEtAl2025,AvkhadievEtAl2026}.

Despite this progress, the extracted long-distance functions remain sensitive
to the adopted functional representation, flavor assumptions, collinear
inputs, kinematic selection, treatment of the transition to the fixed-order
region, and uncertainty construction.  More fundamentally, although
leading-power TMD factorization provides a well-developed organization of
short- and long-distance dynamics and fixes the associated evolution
equations, it does not by itself determine the nonperturbative boundary
information or a unique allocation of the fitted long-distance dependence
among $x$, $b_T$ (or $k_T$), and $Q$.  Nor does it uniquely prescribe the
continuation into the region where $q_T$ becomes comparable to $Q$, for which
matching, power corrections, and the precise observable definition become
essential.  In practice, assumptions about where scale dependence resides, how
the resummed and fixed-order descriptions are connected, and which structural
constraints are imposed on the nonperturbative model enter the definition of
the extracted object.  Because the perturbative kernel, matching prescription,
and functional constraints can materially affect both the central TMD and its
uncertainty, the individual ingredients cannot be interpreted independently
of the composition map that connects them to the physical observable.  Each
component must therefore be specified together with its domain of validity,
scheme, scale dependence, role, and uncertainty, while the composition map
itself must be treated as part of the construction and interpretation of the
result.  Well-defined baseline extractions and controlled comparisons are
essential for determining which features are required by the data and which
are inherited from the chosen representation, matching prescription, or
composition.

Deep neural networks provide a highly expressive alternative to
low-dimensional analytic ansätze, allowing correlated multidimensional TMD
structure to be inferred from data with substantially less restriction on its
functional form.  DNN-based TMD extraction was introduced in our earlier
analysis of the Sivers function, which, to our knowledge, provided the first
application of a deep neural network to the extraction of a TMD
\cite{FernandoKellerSivers2023}.  Neural-network methods were subsequently
applied by the MAP Collaboration to the unpolarized sector, using a
comparatively shallow neural parametrization for the nonperturbative
contribution within a resummed $b_T$-space framework
\cite{BacchettaNN2025}.  Our later work provided, to our knowledge, the first
deep-neural-network extraction of the unpolarized TMD itself, formulated
directly in native $k_T$ space and embedding a jointly $x$-, $Q$-, and
$k_T$-dependent learned profile within an end-to-end differentiable
momentum-space convolution \cite{FernandoKeller2025}.  The distinction is not
merely architectural: representational depth and multidimensional conditioning
change the class of functions that can be inferred from the data and,
consequently, the model dependence exposed by the extraction.  Neural-network
parametrizations of the $b_T$-space nonperturbative sector have most recently
been extended to a joint DY--SIDIS analysis at $\NthreeLL$ accuracy
\cite{Cerutti2026}.

How best to harness this flexibility for phenomenological extraction while
preserving a transparent separation of physical mechanisms remains an active
methodological question.  A fully unconstrained, end-to-end surrogate for the
cross section, for example, would risk obscuring the distinction between
perturbatively calculable QCD contributions and genuine long-distance dynamics.
This separation need not be sacrificed, however, to obtain the advantages of a
direct momentum-space extraction.  Although the perturbative calculation
determines a substantial part of the low-$k_T$ behavior, a DNN formulated
directly in $k_T$ space may provide a more direct means of constraining the
remaining nonperturbative structure in this region, since it is learned in its
native momentum-space representation without an intervening transformation
from $b_T$ space \cite{FernandoKeller2025}.  The impact-parameter-space
extraction developed here is deliberately hybrid: we retain the perturbative
$W$ kernel explicitly, evaluate it at the data kinematics, and use a compact,
physics-informed neural network only for the residual nonperturbative factor.
Positivity, normalization at $\bt=0$, and monotonic damping are built directly
into the architecture.  The resulting model can learn nontrivial $x$--$\bt$
correlations while preserving a transparent separation between perturbative
and nonperturbative contributions.

This work is complementary to the direct momentum-space neural extraction of Ref.~\cite{FernandoKeller2025}, which addresses a different inverse problem by reconstructing the transverse-momentum distribution directly in $\kt$ space without imposing a perturbatively resummed evolution kernel. This strategy has important advantages: observables such as the transverse-momentum width are obtained as normalized moments of the reconstructed momentum-space profile, so that the pronounced $x$ dependence and the milder, systematic $Q$ dependence emerge directly from the measured spectra rather than being prescribed by an evolution model. At the same time, the extracted $\kt$ profile is necessarily an effective one, since perturbative radiation, nonperturbative transverse structure, and evolution effects are not separated within the fit. An ideal comparison would therefore repeat the extraction entirely in $\kt$ space while fixing the short-distance matching coefficients, anomalous dimensions, and $Q$ evolution according to the standard Collins factorization framework. However, no comparably complete $\kt$-space implementation of this framework currently exists at the required logarithmic accuracy; standard resummed TMD analyses are consequently formulated in $\bt$ space. In the present work, we therefore perform a fit in $\bt$ space, where the resummed $W$-term evolution is fixed at $\NthreeLL$ and the hard/OPE matching is retained at strict NLO, while the neural-network component is restricted to the nonperturbative factor. The two approaches thus probe complementary balances between perturbative prior information and data-driven flexibility.

Building on our first use of feature-wise linear modulation (FiLM) in a TMD
extraction \cite{FernandoKeller2025}, we adopt the same conditioning strategy
here \cite{FiLM}.  FiLM is particularly well suited to multidimensional TMD
inference because the longitudinal variable $x$ can condition the hidden
representation of the transverse coordinate $b_T$ through learned
feature-wise scale and shift transformations.  This construction permits
smooth, nonfactorized $x$--$b_T$ dependence within a compact architecture,
while leaving the collinear PDFs, perturbative evolution, matching
coefficients, and Fourier--Bessel transform outside the trainable model.

The present analysis includes the following principal elements:
\begin{enumerate}[leftmargin=1.5em]
  \item a perturbative implementation with $\NthreeLL$ resummed evolution and
  strict-NLO hard/OPE matching in the $W$ term;
  \item general-scale OPE matching and a smooth small-$\bt$ profile tested by
  independent numerical comparisons;
  \item explicit treatment of normalization and point-to-point uncertainties
  for fixed-target and collider data;
  \item $p\bar p$ and $pp$ collider observables, including electroweak
  $\gamma^*/Z$ weights, Tevatron bin conventions, and an LHCb fiducial
  acceptance correction derived with DYTurbo;
  \item 96 converged, independently initialized fits and 50 pseudo-data residual
  fields combined into a $96\times50$ ensemble;
  \item separate estimates of experimental, initialization, combined, and
  collinear-PDF uncertainties;
  \item a controlled extension from the strict low-$q_T/Q$ sample to the
  nominal $q_T/Q\leq0.20$ data set;
  \item a finite-$Y$ robustness test on 24 additional Tevatron points reaching
  $q_T/Q\simeq0.30$, with the nominal 329-point fit held fixed; and
  \item a direct Tevatron $W+Y$ benchmark together with a separate candidate
  refit used to test the identifiability of the nonperturbative factor when the
  specified finite-$Y$ input is supplied.
\end{enumerate}

The paper is organized as follows.  Sec.~\ref{sec:formalism} defines the $b_T$-space factorization and perturbative accuracy.  Sec.~\ref{sec:data} describes the data selection and uncertainty treatment.  Sec.~\ref{sec:method} presents the FiLM network and fitting objective.  Sec.~\ref{sec:validation} reports the perturbative, numerical, and statistical validation tests.  Sec.~\ref{sec:uncertainties} explains the experimental, initialization, combined, and PDF uncertainty constructions.  Results in $b_T$ space and at the cross-section level are given in Sec.~\ref{sec:results}.  Factorization-validity, fixed-order benchmarking, the finite-$Y$ boundary test,
and the exploratory $W+Y$ identifiability study are discussed in
Sec.~\ref{sec:systematics}.  We summarize the scope and limitations in Sec.~\ref{sec:discussion} and conclude in Sec.~\ref{sec:conclusion}.

\section{$\NthreeLL$ $W$-term factorization and matching}
\label{sec:formalism}

\subsection{Drell--Yan structure function}

For dilepton invariant mass $Q$, rapidity $y$, center-of-mass energy $\sqrt{s}$, and transverse momentum $\bm q_T$, the leading-power low-$\qT$ cross section is written schematically as
\begin{equation}
\frac{d\sigma}{dQ^2\,dy\,d^2\bm q_T}
=
\sigma_0(Q,s)\left[W(Q,y,\bm q_T)+Y(Q,y,\bm q_T)\right],
\label{eq:crosssection}
\end{equation}
where $\sigma_0$ contains the electroweak prefactor.  The nominal 329-point fit uses the resummed $W$ contribution, with the finite-$Y$ term set to zero.  Section~\ref{sec:finite-y-boundary} separately examines a smooth finite-$Y$
transition on 24 additional Tevatron points near the upper boundary of the
studied kinematics.  Section~\ref{sec:wy-candidate} then reports an isolated
$W+Y$ candidate constructed from external $W$ and finite-$Y$ inputs on the nominal core.
Both studies are diagnostics and leave the nominal fit and data set unchanged.
Four higher-$q_T$ LHCb points are excluded from these extensions.  The singular
term is
\begin{align}
&W(Q,y,\bm q_T)
=
H(Q,\mu_Q)
\sum_q e_q^2
\int\frac{d^2\bm b_T}{(2\pi)^2}\,
e^{i\bm q_T\cdot\bm b_T}
\nonumber\\
&\times
\widetilde f_{q/h_1}(x_1,\bt;\mu_Q,\zeta_1)
\widetilde f_{\bar q/h_2}(x_2,\bt;\mu_Q,\zeta_2)
+(q\leftrightarrow\bar q),
\label{eq:W}
\end{align}
with $x_{1,2}\simeq (Q/\sqrt{s})e^{\pm y}$ at leading power.  After azimuthal integration, the transform reduces to an integral with $J_0(\qT\bt)$.

The single-TMD quantity reported in this work is
\begin{align}
&\widetilde f_{q/h}(x,\bt;Q)\nonumber\\
&=
\left[C_{q\leftarrow j}\otimes f_{j/h}\right](x;\mub,\zetaB)
e^{-S(\bt,Q)/2}
\FNP(x,\bt),
\label{eq:tmddefinition}
\end{align}
where the hard factor is excluded from the single-TMD definition and is restored at the cross-section level.  Flavor dependence enters through the collinear PDFs and matching coefficients.  The baseline model uses a common $\FNP$ for all light flavors.

\subsection{Evolution, profiles, and logarithmic accuracy}

The perturbative calculation is organized around the resummed
impact-parameter-space \(W\) term.  For each measured point it constructs a
\(b_T\)-space kernel, performs the Fourier--Bessel transform, and multiplies the
result by the fitted nonperturbative factor.  The evolution from the natural \(b\)-space scales to
  the hard scale is written schematically as
  \begin{align}
  &\widetilde f_{q/h}^{\rm pert}(x,b_T;Q)\nonumber\\
  &=
  \left[
  C_{q\leftarrow j}(x,b_T;\mu_b,\zeta_b)
  \otimes f_{j/h}(x,\mu_b)
  \right]\,
  \exp\!\left[-\frac{1}{2}S(b_T,Q)\right],
  \end{align}
  so that the fitted object is
  \begin{equation}
  \widetilde f_{q/h}(x,b_T;Q)
  =
  \widetilde f_{q/h}^{\rm pert}(x,b_T;Q)\,
  F_{\rm NP}(x,b_T).
  \end{equation}
  The hard factor is not included in the reported single-TMD quantity; it is applied
  at the cross-section level in the \(W\) term.

  The scale \(\mu_b\) is generated from a smooth profile.  In the perturbative
  region it follows the canonical inverse-\(b_T\) scale, at very small \(b_T\) it
  is capped by the hard scale \(Q\), and at large \(b_T\) it freezes to avoid
  evaluating the perturbative ingredients below the intended range.  This profile
  is used in the Sudakov evolution and in the collinear PDFs.  The OPE matching
  coefficients are evaluated with a separate smooth perturbative coordinate,
  denoted here by \(b_{\rm pert}\), which agrees with the canonical profile in
  the ordinary perturbative window but avoids spurious large matching logarithms
  when the small-\(b_T\) cap \(\mu_b\to Q\) becomes active.  This is a fixed
  theory-profile prescription, not a fitted degree of freedom.

The quoted logarithmic accuracy refers to the evolution kernel of the
\(W\) term.  The nominal fit uses \(\NthreeLL\) evolution together with explicit
NLO hard and OPE matching inside \(W\).  The NLO organization is strict rather than
  multiplicative: after expanding to first order in \(\alpha_s\), the perturbative
  luminosity entering \(W\) is schematically
  \begin{equation}
  {\cal L}^{W}_{\rm NLO}
  =
  {\cal L}^{(0)}
  +
  \delta H^{(1)}\,{\cal L}^{(0)}
  +
  \delta{\cal L}^{(1)}_{qq}
  +
  \delta{\cal L}^{(1)}_{qg}
  +
  {\cal O}(\alpha_s^2).
  \end{equation}
  Here \({\cal L}^{(0)}\) is the Born quark--antiquark luminosity,
  \(\delta H^{(1)}\) is the one-loop Drell--Yan hard correction, and
  \(\delta{\cal L}^{(1)}_{qq}\) and \(\delta{\cal L}^{(1)}_{qg}\) are the NLO
  quark and gluon OPE matching contributions from the two TMD legs.  Products
  such as \(\delta H^{(1)}\delta C^{(1)}\) and
  \(\delta C^{(1)}\delta C^{(1)}\) are not retained in this strict-NLO
  organization, since they are formally of order \(\alpha_s^2\).

  For the nominal data set the fitted cross sections are therefore
  described as a resummed \(W\)-term calculation with strict-NLO hard/OPE
  matching in \(W\), multiplied by the learned \(F_{\rm NP}\) and by the profiled
  data-set normalizations.  Here and below, the \(\NthreeLL\) designation refers
  to the evolution of the \(W\) term together with the stated strict-NLO hard and
  OPE matching.  It does not claim \(\NthreeLL\) accuracy for the finite-\(Y\)
  transition or for the complete high-\(q_T\) matched cross section.  This
accuracy statement is restricted to the low-\(q_T\), \(W\)-dominated
region used in the nominal fit.  Points outside the nominal low-\(q_T/Q\) range
do not constrain that fit; the 24 additional Tevatron points enter only the
finite-\(Y\) comparison of Sec.~\ref{sec:finite-y-boundary}.

\subsection{NLO hard and OPE insertion}

The hard and OPE matching corrections are inserted in the resummed
$W$ term with a fixed perturbative ordering.  At the cross-section level
the Drell--Yan hard function multiplies the product of the two matched
TMD legs.  If all one-loop factors were left in multiplicative form, the
result would also contain products of one-loop terms, such as
$\delta H^{(1)}\delta C^{(1)}$ and
$\delta C^{(1)}\delta C^{(1)}$.  These products are formally
$\mathcal O(\as^2)$ but would not represent a complete NNLO matching
calculation.  We therefore expand the hard-plus-OPE contribution and
retain only the terms through strict NLO.

Writing the perturbative luminosity relative to its Born value, the
implemented structure is
\begin{equation}
\frac{W_{\rm pert}}{W_{\rm Born}}
=
1+\delta_{\rm hard}+\delta_{qq}+\delta_{qg}
+\mathcal O(\as^2),
\label{eq:strict-nlo-w}
\end{equation}
where $\delta_{\rm hard}$ is the one-loop Drell--Yan hard correction,
$\delta_{qq}$ is the sum of the quark-channel one-loop OPE corrections
from the two TMD legs, and $\delta_{qg}$ is the corresponding gluon-initiated
matching contribution.  Equivalently, the perturbative luminosity entering
$W$ is organized as a Born luminosity plus one insertion of either the hard
correction or an OPE correction.  No product of two NLO insertions is
retained.

The matching coefficients are evaluated with the same general-scale profile
used in the perturbative calculation.  This matters most in the small-$\bt$
transition region, where the scale profile caps $\mu_b$ near the hard scale
and the OPE logarithms must remain consistent with that noncanonical scale
choice.  The smooth $b_{\rm pert}$ coordinate used for the OPE avoids a
spurious large-logarithm behavior at very small $\bt$ while preserving the
canonical behavior in the ordinary perturbative window.  This is part of the
fixed theory prescription; it is not a trainable component of the fit.

At representative fixed-target points, the NLO contributions are
\begin{align}
\delta_{qq}&\in[-0.0781,-0.0682],&
\delta_{qg}&\in[0.0042,0.0203],\nonumber\\
\delta_{\rm OPE}&\in[-0.0720,-0.0516],&
\delta_{\rm hard}&\in[0.1201,0.1424].
\end{align}
Here $\delta_{\rm OPE}=\delta_{qq}+\delta_{qg}$, and the quoted
$\delta_{\rm OPE}$ and net $\delta_{\rm hard}+\delta_{\rm OPE}$ intervals are
the extrema of point-by-point sums across the sampled kinematics, not sums of
independently selected interval endpoints.  The pointwise strict correction
ranges from $5.2\%$ to $9.1\%$ at these canonical points.  The largest
difference between the unexpanded multiplicative product and the strict result
in the corresponding comparison is approximately $1.3\times10^{-3}$ relative
to the Born contribution.
This confirms that the omitted product terms are numerically small in the
validated low-$q_T$ region, but the strict expansion is retained so that the
perturbative accuracy statement remains unambiguous.

The accuracy label assigned in this work therefore refers to the resummed
$W$ term with strict-NLO hard/OPE matching.  The finite-$Y$ term and the
large-$q_T$ fixed-order tail are treated separately in
Sec.~\ref{sec:systematics}; the reported single-TMD quantity is not a
fixed-order high-$k_T$ prediction.

\subsection{Regularized momentum-space companion}

The primary object extracted in this analysis is
$\widetilde f_{q/h}(x,\bt;Q)$ in impact-parameter space.  This is the natural
space for the resummed evolution, the OPE matching, and the neural
nonperturbative factor.  For comparison with more familiar momentum-space
representations we also construct the companion distribution
\begin{equation}
f_{q/h}(x,\kt;Q)
=
\frac{1}{2\pi}\int_0^\infty d\bt\,\bt\,
J_0(\kt\bt)\,
\widetilde f_{q/h}(x,\bt;Q).
\label{eq:hankel}
\end{equation}
Equation~\eqref{eq:hankel} is a Bessel transform of the fitted $b_T$-space TMD,
not an independent fit in $\kt$ space.  Its numerical evaluation requires an
explicit prescription because the fitted curves are known only on a finite
$\bt$ grid.  We first interpolate each $b_T$-space curve with a shape-preserving
interpolator.  Beyond the end of the fitted grid, the curve is continued with a
smooth large-$\bt$ damping form matched to its final segment, and the end of
the extended integration domain is multiplied by a smooth taper.  The same
prescription is applied to every member of the uncertainty ensemble.

The regularization has two consequences for interpretation.  First, the
low- and moderate-$\kt$ behavior shown in the paper is a faithful transform of
the fitted $b_T$-space result over the region where the inverse-transform closure
tests pass.  Second, the companion should not be interpreted as a fixed-order
large-$\kt$ tail.  At large transverse momentum the correct prediction would
require the fixed-order matching structure of the cross section, not only the
finite-grid transform of the nonperturbative TMD in $b_T$ space.

\section{Data and uncertainty model}
\label{sec:data}

\subsection{Data selection}\label{sec:data-selection}

The analysis begins with a fixed-target DY sample built from E288, E605, and
E772 \cite{E288,E605,E772}.  The E288 measurements are separated by beam
energy and denoted E288-200, E288-300, and E288-400.  This sample contains 418
points.  Evaluated with its corresponding uncertainty prescription, the mean
squared standardized residual is 0.958; the largest data-set-level value is
1.43 and the largest absolute normalization pull is 0.230 standard deviations.

The nominal data set adds the collider measurements for which the observable
definition, binning, normalization, and uncertainty treatment can be implemented
consistently.  These are the absolute CDF Run I, CDF Run II, and D0 Run I
$p\bar p\to Z/\gamma^*+X$ spectra in the $Z$ region
\cite{CDFRunI,CDFRunII,D0RunI}, together with the forward 7 TeV
$pp\to Z/\gamma^*+X$ spectrum from LHCb \cite{LHCb7Z}.  The LHCb prediction is
obtained from the boson-level calculation multiplied by a DYTurbo-derived
fiducial-to-inclusive acceptance factor \cite{DYTurbo,CamardaN3LL}.  The same
mass, rapidity, and $q_T$ bins are used in the acceptance calculation.  The
fiducial convention is checked against DYTurbo and MCFM
\cite{DYTurbo,CamardaN3LL,CuTeMCFM}; the MCFM result is divided by two because its symmetric $|\eta|$ cuts include
both positive- and negative-rapidity hemispheres, whereas the LHCb measurement
uses only the positive-rapidity hemisphere.

A strict low-transverse-momentum sample is defined by
\begin{equation}
q_T/Q \le 0.10 .
\end{equation}
It contains 180 points and has a mean squared standardized residual of 0.521
under its corresponding uncertainty prescription.  This sample is used to test
the stability of the extraction as the kinematic range is enlarged.  The
factorization-validity uncertainty of Sec.~\ref{sec:factorization-validity}
allows fixed-target and Tevatron points through $q_T/Q\le0.20$, while the LHCb
sample remains restricted to its low-$q_T/Q$ subset.  The resulting nominal
sample contains $N_{\rm acc}=329$ points.  Its unpenalized data contribution is
\begin{equation}
\begin{aligned}
\frac{\chi^2_{\rm data}}{N_{\rm acc}}&\simeq0.418,
& N_{\rm acc}&=329,\\
\chi^2_{\rm data}&\simeq137.46.
\end{aligned}
\label{eq:nominal-data-chi2-summary}
\end{equation}
The standardized residuals use the effective uncertainty of
Eq.~\eqref{eq:sigma-eff-factorization} and are evaluated after profiling the
data-set normalizations.  Equation~\eqref{eq:nominal-data-chi2-summary} does not
include the normalization or empirical-reference contributions to the fitting
objective.  Because the strict and nominal samples use different effective
uncertainties, their fit-quality values should not be interpreted as a nested
sequence evaluated with an identical statistical model.

\begin{table*}[t]
\caption{Nominal 329-point data set.  The global unpenalized data
contribution is $\chi^2_{\rm data}/N_{\rm acc}\simeq0.418$
($\chi^2_{\rm data}\simeq137.46$) under the effective uncertainty model of
Sec.~\ref{sec:factorization-validity}.  The normalization and empirical-reference
terms are specified separately in Sec.~\ref{sec:fit-quality}.  In the final
column, ``diag.'' denotes diagonal point-to-point uncertainties, ``p2p'' an
additional point-to-point component, ``norm.'' a normalization prior,
``fact.'' the factorization-validity contribution, and ``fid. acc.'' the
fiducial-acceptance treatment.}
\label{tab:data}
\begin{ruledtabular}
\begin{tabular}{lccccl}
Dataset & $N$ & $Q$ [GeV] & $\sqrt{s}$ [GeV] & med. $T/D$ & uncertainty treatment\\
\hline
E288-200 & 31 & 4.5--8.5   & 19.37 & 0.989 & diag. + 15\% norm. + fact.\\
E288-300 & 41 & 4.5--11.5  & 23.72 & 0.997 & diag. + 15\% norm. + fact.\\
E288-400 & 63 & 5.5--13.5  & 27.39 & 0.999 & diag. + 5\% p2p + 15\% norm. + fact.\\
E605     & 54 & 7.5--15.75 & 38.8  & 0.978 & diag. + p2p + 15\% norm. + fact.\\
E772     & 54 & 5.5--14.5  & 38.8  & 0.967 & diag. + 5\% p2p + 15\% norm. + fact.\\
CDF Run I  & 30 & 91 & 1800 & 0.958 & diag. + 3.9\% norm. + fact.\\
CDF Run II & 36 & 91 & 1960 & 0.936 & diag. + 5.8\% norm. + fact.\\
D0 Run I   & 14 & 90 & 1800 & 0.979 & diag. + 4.4\% norm. + fact.\\
LHCb 7 TeV & 6  & 90 & 7000 & 1.021 & diag. + fid. acc.\\
\hline
Total & 329 & 4.5--91 & 19.37--7000 & 0.984 & nominal $q_T/Q\le0.20$ sample\\
\end{tabular}
\end{ruledtabular}
\end{table*}

\subsection{Data conversions and collider-specific treatment}

The published fixed-target cross sections were converted to a common differential convention consistent with the theoretical calculation, including the tabulated kinematic prefactors and bin definitions.  The resulting tables were checked for continuity across neighboring kinematic bins.

For the Tevatron $p\bar p$ measurements, the calculation uses antiproton parton luminosities, includes $\gamma/Z$ neutral-current weights over the published mass windows, applies the $p_T$-bin Jacobian for the pb/GeV convention, and applies the rapidity-inclusive scaling needed for the published inclusive $p_T$ spectra.

For LHCb 7 TeV the published fiducial spectrum is compared with a boson-level prediction multiplied by a DYTurbo fiducial/inclusive acceptance factor.  The factor uses the lepton cuts $p_T^\ell>20~\GeV$ and $2<\eta_\ell<4.5$ for both leptons, with matching mass, rapidity, and $q_T$ bins.  Only the low-$q_T/Q$ subset specified above is included.  In the external fixed-order benchmark, the corresponding MCFM fiducial calculation is scaled by $1/2$ before comparison because the implemented MCFM $|\eta|$ cuts include both positive- and negative-rapidity hemispheres, whereas the LHCb data use only the positive-rapidity hemisphere.

\subsection{Normalization, covariance status, and exclusions}

The baseline uses explicit dataset normalization nuisances where the experimental information supports them.  Fixed-target samples use 15\% normalization priors.  The Tevatron normalizations are 3.9\% for CDF Run I, 5.8\% for CDF Run II, and 4.4\% for D0 Run I.  The LHCb 7 TeV points are included without a separate normalization nuisance in the nominal fit.

Where a usable covariance matrix or correlated-systematics model is not available locally, the published tabulated uncertainties are used as diagonal point-to-point errors, as is common in phenomenological fits.  This is a pragmatic data-use prescription, not a claim that correlations are absent.  CDF Run II is the most important limitation: the publication describes correlated efficiency/systematic structure, but the numerical tables used here do not provide a complete covariance representation for the present fit.  The present fit therefore uses the published combined table uncertainty as diagonal plus the published normalization nuisance.

The normalized D0 Run II spectra are not included because the present
calculation is formulated for absolute rather than normalized distributions.
Twenty-four additional Tevatron points are used only in the finite-$Y$ study of
Sec.~\ref{sec:finite-y-boundary} and are not part of the nominal 329-point
sample.  Four higher-$q_T$ LHCb points are excluded because a consistent
observable and covariance treatment is not available within the present
analysis.

\section{Physics-informed neural extraction}
\label{sec:method}

\subsection{Prediction model}

For each measured point, the perturbative calculation is evaluated before
neural-network optimization and represented by a $W$-term kernel $K_{i\ell}$ on
the $b_T$ grid.  The nominal prediction is
\begin{equation}
T_i^{(W)}(\theta)
=
\sum_\ell K_{i\ell}
\FNP(x_{1i},b_\ell;\theta)
\FNP(x_{2i},b_\ell;\theta).
\label{eq:prediction}
\end{equation}
The kernel contains the Bessel weight, integration measure, strict-NLO matched
perturbative luminosity, target/isospin treatment, and cross-section convention.
A direct finite-$Y$ contribution is not included in Eq.~\eqref{eq:prediction}.
The separate transition of Sec.~\ref{sec:finite-y-boundary} is evaluated only for
the 24 additional Tevatron points and does not modify the nominal fit.  A
data-set normalization nuisance $n_{d(i)}$ gives
\begin{equation}
\widehat T_i=n_{d(i)}T_i^{(W)}.
\end{equation}

The data and normalization contributions are defined by the following expressions,
with $t_i$ denoting the measured cross section in the central fit:
\begin{align}
\chi^2_{\rm data}(\theta,n)
&=
\sum_{i=1}^{N_{\rm acc}}
\left[
\frac{\widehat T_i(\theta,n)-t_i}{\sigma_{i,\rm eff}}
\right]^2,
\label{eq:chi2-data}\\
\chi^2_{\rm norm}(n)
&=
\sum_d
\left[
\frac{n_d-1}{\Delta_d}
\right]^2.
\label{eq:chi2-norm}
\end{align}
The effective point uncertainty $\sigma_{i,\rm eff}$ is defined in
Sec.~\ref{sec:factorization-validity}; $\Delta_d$ is the normalization prior for
data set $d$.

To reduce optimization-induced nonuniqueness without imposing a fixed functional
form, the nominal objective includes a pointwise distance in $\FNP$ itself,
rather than in $\log\FNP$.  The reference curve $\FNP^{\rm ref}$ is the pointwise median of an ensemble of
fits obtained from independent parameter perturbations under the same architecture
and data selection.  The reference distance is evaluated on
\begin{equation}
\mathcal X_{\rm ref}
=
\{0.001,0.003,0.01,0.03,0.1,0.2,0.4,0.7\}
\label{eq:reference-x-grid}
\end{equation}
and on the fixed numerical set of $b_T$ nodes
\begin{equation}
\mathcal B_{\rm ref}
\equiv
\{b_a\}_{a=1}^{N_b^{\rm ref}},
\qquad
0.1\le b_a\le2.0~\GeV^{-1}.
\label{eq:reference-b-domain}
\end{equation}
Here $N_b^{\rm ref}=|\mathcal B_{\rm ref}|$, and the same set
$\mathcal B_{\rm ref}$ is used for every fit.  At each point
$(x_j,b_a)\in\mathcal X_{\rm ref}\times\mathcal B_{\rm ref}$ we define
\begin{equation}
r^{\rm ref}_{ja}(\theta)
=
\frac{
\FNP(x_j,b_a;\theta)-\FNP^{\rm ref}(x_j,b_a)
}{
\max[\FNP^{\rm ref}(x_j,b_a),0.10]
}.
\label{eq:reference-residual}
\end{equation}
For $N_{\rm ref}=|\mathcal X_{\rm ref}|\,N_b^{\rm ref}$, the
reference contribution is normalized to the number of nominal data points,
\begin{equation}
\chi^2_{\rm ref}(\theta)
=
\frac{N_{\rm acc}}{N_{\rm ref}}
\sum_{j,a}
\left[r^{\rm ref}_{ja}(\theta)\right]^2.
\label{eq:chi2-reference}
\end{equation}
The dimensionless coefficient $\lambda_{\rm ref}$ controls the relative weight
of this term in the total objective,
\begin{align}
\Phi(\theta,n)
&=
\chi^2_{\rm data}
+
\chi^2_{\rm norm}
+
\lambda_{\rm ref}\chi^2_{\rm ref},
\label{eq:objective-decomposition}\\
\mathcal L(\theta,n)
&=
\frac{\Phi(\theta,n)}{N_{\rm acc}},
\qquad
\lambda_{\rm ref}=1.
\label{eq:loss}
\end{align}
Thus the data-point-normalized reference term enters with unit coefficient.
No additional penalty on the functional path length, on $\log F_{\rm NP}$, on curvature, or on the large-$b_T$ tail is included in the nominal objective.

The reference term is a soft regularization over the ranges in
Eqs.~\eqref{eq:reference-x-grid} and \eqref{eq:reference-b-domain}.  It neither
fixes $\FNP(x,2~\GeV^{-1})$ nor imposes a unique mathematical solution.  The
same empirical reference curve is used for all fits rather than being recomputed
after excluding each fit in turn.  The resulting ensemble is therefore
conditional on this common reference construction.

\subsection{FiLM-conditioned monotone architecture}

Rather than fitting $\FNP$ directly, the network learns a nonnegative damping rate $A_\theta(x,\bt)$.  The radial inputs are expressed through the dimensionless coordinate
\begin{equation}
\bar b_T \equiv \frac{b_T}{1~\GeV^{-1}},
\end{equation}
and the radial features are
\begin{equation}
\psi(\bar b_T)
=
\left(
\bar b_T,\bar b_T^2,\sqrt{\bar b_T+\epsilon},\ln(1+\bar b_T)
\right),
\qquad \epsilon=10^{-8}.
\end{equation}
The numerical implementation therefore evaluates these features using the value of $b_T$ in $\GeV^{-1}$.  The conditioning features are
\begin{equation}
c(x)=
\left(
x,\ln\frac{x}{1-x}
\right).
\end{equation}
The radial lift is a $4\to48$ dense layer with a hyperbolic-tangent activation.  The conditioning stream is a $2\to32\to32$ multilayer perceptron with SiLU activations.  Three residual FiLM blocks modulate the 48-dimensional radial representation with scale and shift vectors generated by the $x$ stream \cite{FiLM}:
\begin{align}
u^{(\ell)}&=\phi\!\left(W_1^{(\ell)}h^{(\ell)}\right),\\
\widetilde u^{(\ell)}
&=
\gamma^{(\ell)}(c)\odot u^{(\ell)}
+\beta^{(\ell)}(c),\\
h^{(\ell+1)}
&=
\phi\!\left(W_2^{(\ell)}\widetilde u^{(\ell)}
+h^{(\ell)}\right).
\end{align}
The final head is
\begin{equation}
A_\theta(x,\bt)
=
\mathrm{softplus}
\left(
w_{\rm out}^{T}h^{(3)}+b_{\rm out}
\right)+a_{\min},
\label{eq:rate}
\end{equation}
with $a_{\min}=0$ in the nominal fit.  The output $A_\theta$ is interpreted in $\GeV^2$, so the exponent below is dimensionless, and $A_\theta\ge0$.  The physical nonperturbative factor is constructed as
\begin{align}
I_\theta(x,\bt)
&=
\int_0^{\bt} d\beta\,2\beta A_\theta(x,\beta),\\
\FNP(x,\bt)
&=
\exp[-I_\theta(x,\bt)].
\label{eq:FNP}
\end{align}
Equations~\eqref{eq:rate} and \eqref{eq:FNP} guarantee
\begin{equation}
\FNP(x,0)=1,\qquad
\frac{\partial\FNP}{\partial\bt}\le0.
\end{equation}

\begin{figure*}[t]
\centering
\paperfigure{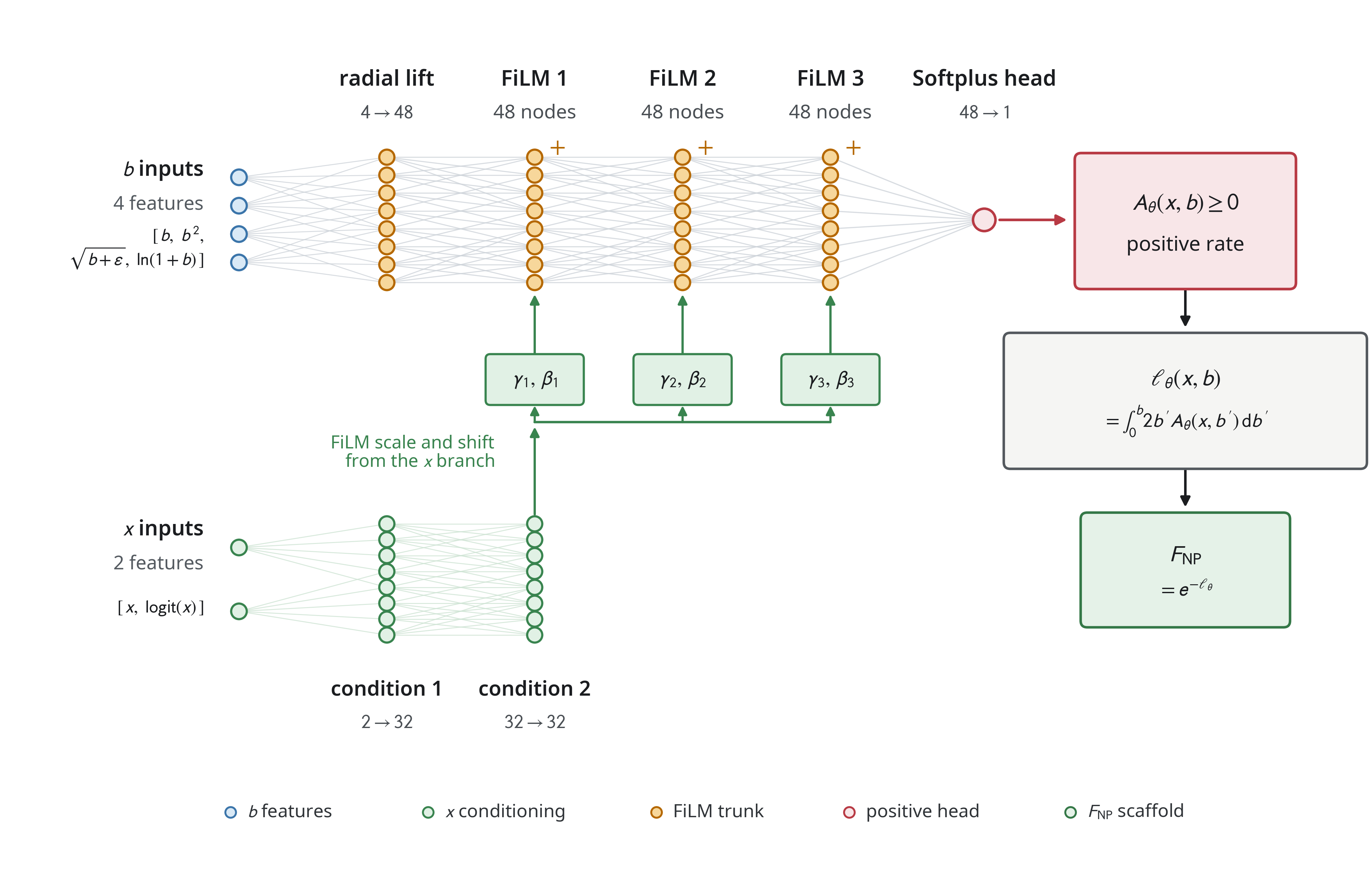}{0.96}
\caption{Node-level schematic of the FiLM-conditioned network used for the nonperturbative factor.  The radial branch lifts four smooth dimensionless $\bt$ features into a 48-dimensional residual trunk.  A two-feature $x$ branch generates FiLM scale and shift vectors for three residual blocks.  A Softplus head produces the positive damping rate $A_\theta(x,\bt)$, which is integrated to form the monotone factor $\FNP(x,\bt)$.  The perturbative $\NthreeLL$ $W$ kernel is evaluated outside the network.}
\label{fig:architecture}
\end{figure*}

\subsection{Fitting strategy}

Figure~\ref{fig:architecture} illustrates the restricted role of the neural
network.  The $b_T$ branch controls the radial shape, while the $x$ branch
supplies the FiLM scale and shift parameters that allow the damping rate to vary
smoothly with partonic momentum fraction.  The Softplus head and subsequent
integral turn an unconstrained network output into a positive rate and then into
a monotone factor with the required normalization at $b_T=0$.  The collinear
PDFs, hard coefficient, OPE matching, perturbative evolution, and
Fourier--Bessel integration remain outside the trainable model.

The architecture is intentionally less general than an end-to-end cross-section
network.  It can learn smooth, nonfactorized correlations between $x$ and $b_T$
within the support of the data, but it cannot alter the fixed perturbative
components or generate a sign-changing nonperturbative factor.  The extracted
$F_{\rm NP}$ is therefore defined relative to the perturbative framework adopted
in this analysis.

The inverse problem does not select a unique optimized network.  We therefore
perform 96 independent optimizations from perturbed versions of a common
initialization and retain only solutions satisfying the convergence criterion in
Sec.~\ref{sec:initialization-stability}.  Experimental variation is represented
by 50 residual fields obtained from pseudo-data fits, as defined in
Sec.~\ref{sec:validation}.  Combining these two components produces $4{,}800$
members per flavor, but only 96 independent neural-network optimizations.  The
quoted uncertainty is consequently conditional on the FiLM architecture, the
monotone construction, the empirical-reference objective, the specified
initialization family, and the procedure used to combine the two uncertainty
components.

The condition $\FNP(x,0)=1$ is structural because it follows from the integrated
rate in Eq.~\eqref{eq:FNP}.  By contrast, the empirical-reference term acts only
over $0.1\le b_T\le2.0~\GeV^{-1}$.  No condition is imposed at
$b_T=2~\GeV^{-1}$.  Behavior outside the reference interval is determined by
the network construction and the data-weighted objective; the momentum-space
representation additionally depends on the finite-$b_T$ continuation described
in Appendix~\ref{app:hankel}.

\section{Validation of the perturbative and statistical implementation}
\label{sec:validation}

A good cross-section fit alone is insufficient because a flexible network could
compensate for a normalization error, an inconsistent OPE convention, or an
unstable scale profile.  We therefore test the perturbative calculation before
fitting, verify the pseudo-data construction independently, and examine the
stability of the fitted TMDs with respect to initialization and ensemble size.

\subsection{Perturbative implementation and internal consistency}

At the benchmark points, independent evaluations of the Born-level parton
luminosity and of its embedding in the Born limit of the $W$ term agree to within
$7\times10^{-16}$ in relative accuracy, consistent with double-precision
roundoff.  At NLO we separately verify the quark--quark and quark--gluon
contributions to the small-$b_T$ OPE, the normalization convention for
$\alpha_s$, and the strict truncation through $\mathcal O(\alpha_s)$.  All
quantities evaluated at the benchmark points are finite.

The strict NLO correction factor, defined as the hard-plus-OPE contribution
truncated through $\mathcal O(\alpha_s)$ divided by its Born value, remains
positive throughout the benchmark set.  To estimate formally higher-order terms
introduced by leaving the NLO factors unexpanded, we compare the strict result
with the corresponding unexpanded product.  Their maximum relative difference
is $8.1\times10^{-3}$, or approximately $0.81\%$.  These additional terms begin
at $\mathcal O(\alpha_s^2)$ but do not constitute a complete prediction at that
order.

\subsection{General-scale OPE and the small-$b_T$ profile}

The perturbative Fourier--Bessel integrand samples regions in which the profile
scales differ from their canonical values.  We therefore retain the general-scale
form of the small-$b_T$ OPE rather than evaluating the matching coefficients only
at the canonical scale.  Particular care is required at very small $b_T$, where
the canonical scale $\mu_b\sim1/b_T$ would exceed the hard scale $Q$.  Without a
profile for the logarithmic coordinate, the strictly truncated NLO correction
can become negative.  With the smooth profile used here,
\begin{align}
\min R_{\rm strict}^{\rm unprofiled} &= -12.62,\\
\min R_{\rm strict}^{\rm profiled} &= 1.076.
\end{align}
The profile removes the negative contribution over the validation range.  The
maximum relative difference between the strict and unexpanded products over the
broader general-scale comparison is approximately $1.08\%$.  The three quoted
strict-versus-unexpanded comparisons refer to different domains: the
$1.3\times10^{-3}$ value in Sec.~\ref{sec:formalism} is Born-normalized at
representative canonical-scale fixed-target points, the $8.1\times10^{-3}$ value
above is the maximum fractional difference over the separate benchmark set, and
$1.08\%$ is the maximum over this broader profiled general-scale scan.

\subsection{Singular subtraction and external comparisons}

The singular contribution entering the additive matching prescription is
implemented in two independent forms.  After matching the perturbative
truncation, normalization, scale choices, and subtraction conventions, the two
calculations agree within numerical precision.  Denoting the fixed-order
cross section by ${\rm FO}$ and its singular small-$q_T$ limit by
${\rm singular}$, the finite matching contribution is defined by
\begin{equation}
Y={\rm FO}-{\rm singular}
\end{equation}
closes numerically at the $10^{-17}$ level over the benchmark set.

The finite-order tail is also compared with independent MCFM and DYTurbo
calculations at representative fixed-target, Tevatron, and LHCb points.  The
comparison uses the same mass windows, rapidity conventions, bin definitions,
and fiducial selections as the processed data.  Agreement verifies the
normalization, unit conversion, and kinematic dependence of the real-emission
contribution at the selected points \cite{DYTurbo,CamardaN3LL,CuTeMCFM}.
Measurements outside the strict low-$q_T/Q$ region are nevertheless included
only with the factorization-validity uncertainty of Sec.~\ref{sec:systematics},
rather than as unweighted leading-power constraints.

\subsection{Central-fit quality}
\label{sec:fit-quality}

For the 418-point fixed-target sample, the mean squared standardized residual is
0.958 under the uncertainty prescription used for that fit.  The largest
data-set-level value is 1.43, and the largest absolute normalization pull is
0.230 standard deviations.  This provides a fixed-target comparison, although it
is not directly comparable to the nominal fit because the data selection and
effective uncertainties differ.

For the nominal 329-point fit, the data, normalization, and empirical-reference
terms are separated according to
Eqs.~\eqref{eq:chi2-data}--\eqref{eq:objective-decomposition}.  The data term is
\begin{equation}
\begin{aligned}
\frac{\chi^2_{\rm data}}{N_{\rm acc}}&\simeq0.418,
& N_{\rm acc}&=329,\\
\chi^2_{\rm data}&\simeq137.46.
\end{aligned}
\label{eq:nominal-data-chi2}
\end{equation}
This quantity is evaluated after profiling the data-set normalizations and using
$\sigma_{i,\rm eff}$, so the factorization-validity contribution enters through
the residual denominator.  It does not include $\chi^2_{\rm norm}$ or
$\lambda_{\rm ref}\chi^2_{\rm ref}$.  Because no effective neural-network
degree-of-freedom count is assigned, we do not identify it with
$\chi^2/{\rm d.o.f.}$ or a reduced $\chi^2$.  The global median
prediction-to-data ratio is 0.984.  The relatively small value of the data term
reflects the normalization priors and conservative factorization-validity
uncertainty and is not, by itself, evidence of calibrated coverage or a complete
uncertainty model.

\subsection{Calibration of the pseudo-data fluctuations}

Let $t_i^{(r)}$ denote the sampled value of data point $i$ in pseudo-data
realization $r$, and define its realization average by
\begin{equation}
\overline{t}_i=
\frac{1}{N_{\rm rep}}
\sum_{r=1}^{N_{\rm rep}}t_i^{(r)}.
\end{equation}
In a separate fixed-target-only calibration ensemble of 50 pseudo-data
realizations, all 418 data points have nonzero sample variance.  The realization
means are centered on the measurements,
\begin{equation}
\frac{1}{N_{\rm dat}}
\sum_{i=1}^{N_{\rm dat}}
\frac{\overline{t}_i-D_i}{\sigma_i}
=-0.007.
\end{equation}
After removing a robust estimate of the data-set-wide normalization displacement,
the median ratio of the realization-level shape spread to the quoted pointwise
uncertainty is 1.038.  This fixed-target-only calibration ensemble therefore
reproduces the supplied point-to-point uncertainties to within approximately
$3.8\%$ in the median and is distinct from the separate set of 50 pseudo-data
fits on the nominal 329-point sample.

For each fit in the nominal 329-point pseudo-data ensemble, the change of every
tabulated TMD quantity relative to the central fit is retained on the same
numerical grids.  We refer to this
pointwise change as an \emph{experimental-residual field}.  These fields are
conditional on the common architecture, empirical reference, perturbative
calculation, and fitting protocol.  Because the pseudo-data are generated with
$\sigma_{i,\rm eff}$, they propagate the adopted effective-error model---the
published pointwise uncertainty together with the factorization-validity
contribution where applicable---rather than a purely experimental error model.
They do not imply an independent fit at every initialization.

\subsection{Stability of the experimental-residual ensemble}

For the separate 50 pseudo-data fits on the nominal 329-point sample used to
construct the experimental-residual fields, let $\overline\chi_r^{\,2}$ denote
the mean squared standardized residual,
evaluated with the effective point uncertainties after profiling the
normalization nuisances.  Its empirical distribution satisfies
\begin{align}
\underset{r}{\operatorname{median}}\!\left(\overline\chi_r^{\,2}\right)&=1.45,\\
q_{0.95}\!\left(\overline\chi_r^{\,2}\right)&=1.60,\\
q_{0.95}\!\left(|p_{{\rm norm},r}|\right)&=2.97,
\end{align}
where $p_{{\rm norm},r}$ is the normalization-pull statistic and $q_p$ denotes an
empirical quantile.  To test the finite size of this 50-member nominal
experimental-residual ensemble, we form 500 random partitions into two subsets
of 25.  For subset medians $m_A(z)$ and
$m_B(z)$, and $q_{16}$--$q_{84}$ half-widths $h_A(z)$ and $h_B(z)$, define
\begin{align}
d_{\rm center}(z)&=
\frac{|m_A(z)-m_B(z)|}{\max[m_{AB}(z),\epsilon_{\rm num}]},
\label{eq:center-stability-definition}\\
d_{\rm width}(z)&=
\frac{|h_A(z)-h_B(z)|}{\max[h_{AB}(z),\epsilon_{\rm num}]},
\label{eq:width-stability-definition}
\end{align}
with $m_{AB}=(|m_A|+|m_B|)/2$ and $h_{AB}=(|h_A|+|h_B|)/2$.
The comparison is restricted to the active region
\begin{equation}
\mathcal A_c=
\left\{z\in c:
|m_{\rm full}(z)|>0.05\max_{z'\in c}|m_{\rm full}(z')|
\right\}.
\label{eq:active-region}
\end{equation}
Taking the 90th percentile over the active points, the 95th percentile over
curves, and then the 90th percentile over random partitions gives
\begin{align}
q_{0.90}^{\rm width}&<0.46,\\
q_{0.90}^{\rm center}&<0.08.
\end{align}
The prespecified criteria are
\begin{equation}
q_{0.90}^{\rm width}<0.80,
\qquad
q_{0.90}^{\rm center}<0.08,
\label{eq:experimental-residual-stability}
\end{equation}
and both are satisfied.  All experimental-residual quantile intervals are finite
and nonzero in the active region.

\subsection{Convergence and dependence on initialization}
\label{sec:initialization-stability}

The initialization-dependent component is represented by 96 independently
optimized solutions.  After a minimum training exposure, $\FNP$ is evaluated on
$\mathcal X_{\rm ref}\times\mathcal B_{\rm ref}$ at the end of successive training blocks.  Let $D_F^{(m)}$
denote the maximum pointwise relative change between blocks, evaluated with the
specified numerical floor.  A solution is included only if
\begin{equation}
D_F^{(m)}\le0.02
\label{eq:convergence-criterion}
\end{equation}
holds for the specified number of consecutive training blocks.  The same
criterion is applied to every independent optimization.

To test stability with respect to ensemble size, the number of independent
starts is increased from 48 to 96.  All 48 additional optimizations satisfy the
same convergence criterion.  Let $w_N$ denote the $q_{16}$--$q_{84}$ width of
the initialization-dependent envelope obtained from $N$ solutions.  The
96-to-48 width ratios are
\begin{align}
\frac{w_{96}^{(b)}}{w_{48}^{(b)}}&=1.014,
&\text{$b_T$ space},\nonumber\\
\frac{w_{96}^{(u,k)}}{w_{48}^{(u,k)}}&=1.019,
&\text{$u$ in $k_T$ space},\nonumber\\
\frac{w_{96}^{(d,k)}}{w_{48}^{(d,k)}}&=1.019,
&\text{$d$ in $k_T$ space}.
\label{eq:initialization-width-ratios}
\end{align}
The near-unity ratios show that the quoted spread is stable under this doubling
of the ensemble within the specified initialization and perturbation family.
They do not establish completeness with respect to other architectures,
objectives, initialization procedures, or disconnected solution branches.

\section{Uncertainty propagation}
\label{sec:uncertainties}

\subsection{Construction of the uncertainty ensembles}
\label{sec:combined-ensemble}

The uncertainty analysis separates sensitivity to optimization from the
propagation of experimental errors.  The initialization component consists of
96 independently optimized solutions satisfying the convergence criterion in
Eq.~\eqref{eq:convergence-criterion}.  The experimental component consists of
the 50 experimental-residual fields defined in Sec.~\ref{sec:validation}.  Each
residual field is combined pointwise with each independently initialized fit,
producing $96\times50=4{,}800$ members per flavor.  This construction contains
96 fitted networks and 50 experimental-residual fields.  The number of independently initialized fits was increased until the
initialization-dependent envelope stabilized; doubling the ensemble from 48 to
96 fits changed the relevant widths by only approximately $1.4$--$1.9\%$,
indicating effective saturation within the specified initialization and
perturbation family.

For a tabulated scalar quantity $g(z)$, the initialization envelope is the
empirical $q_{16}$--$q_{84}$ range over the 96 fitted solutions with the
experimental field set to zero.  The experimental envelope is the corresponding
range over the 50 residual fields applied to the central fitted solution.  The
combined envelope is the empirical $q_{16}$--$q_{84}$ range over all $4{,}800$
combinations.  The experimental-residual envelope is the empirical central
68\% interval, defined by the $q_{16}$ and $q_{84}$ quantiles, of the adopted
effective-error model.  It therefore includes the published pointwise
experimental uncertainty and, where applicable, the factorization-validity
contribution entering $\sigma_{i,\rm eff}$.  The initialization and combined
envelopes instead quantify empirical spread within the specified model and
optimization procedure and are not assigned a calibrated confidence level.

All fitted solutions use the same empirical reference curve.  The quoted
initialization and combined envelopes therefore do not include the additional
variation that would arise if the reference curve were reconstructed after
withholding each independently initialized fit.

\subsection{Experimental-residual fields and normalization nuisances}

For pseudo-data realization $r$ and data set $d$, the Monte Carlo data are
constructed schematically as
\begin{equation}
t_i^{(r)}=
\left(1+\delta_d z_d^{(r)}\right)D_i
+\sigma_{i,\rm unc}z_i^{(r)},
\end{equation}
with independent standard normal variates $z_d$ and $z_i$.  The same
normalization widths are represented by profiled nuisance factors in the fit.
Consequently, a substantial part of the correlated experimental variation is
absorbed by data-set normalizations rather than by the universal $\FNP$.  The
TMD-shape response can therefore be narrower than the observable-space
cross-section spread.

The main $b_T$-space and cross-section bands are the combined envelopes defined
in Sec.~\ref{sec:combined-ensemble}, with collinear-PDF uncertainty excluded.
The experimental-only envelope is obtained from the 50 residual fields with the
initialization component held fixed.  The statistics in Sec.~\ref{sec:validation}
and Appendix~\ref{app:experimental-residuals} characterize this 50-member
experimental component; they do not imply an independent neural-network fit for
every one of the $4{,}800$ combined members.

\subsection{Collinear-PDF dependence}

The 96 neural-network fits use the central NNPDF4.0 member
\cite{NNPDF40,LHAPDF}.  Collinear-PDF variation is evaluated separately from the
$4{,}800$-member combined envelope.  In the first estimate, a representative
experimental member is reconstructed with a noncentral PDF member,
\begin{equation}
\widetilde f_q^{(r,m_r)}=
\left[C\otimes f^{(m_r)}\right]_q
e^{-S^{(m_r)}/2}\FNP^{(r)},
\label{eq:overlay}
\end{equation}
and the resulting 50 experimental--PDF pairs are summarized by their empirical
$q_{16}$ and $q_{84}$ quantiles.  This estimate varies the collinear PDFs in the
TMD reconstruction but does not refit $F_{\rm NP}$ for each PDF member.

A second comparison recomputes the full $W(b_T)$ contribution for selected
noncentral NNPDF4.0 members while holding the fitted $F_{\rm NP}$ fixed.  The
largest effects occur in high-$x$ fixed-target kinematics, where the antiquark
luminosity uncertainty can exceed the uncertainty propagated from the
experimental residuals.  Representative NNPDF4.0 NNLO nominal 68\% half-widths
are 0.152 for E288-200 at $Q=8.5~\GeV$, 0.551 for E288-400 at
$Q=12.5~\GeV$, and 0.385 for E605 at $Q=15.75~\GeV$, compared with
experimental half-widths of 0.0216, 0.0129, and 0.0084 in the same panels.  The
PDF dependence is therefore displayed separately rather than combined with the
initialization and experimental envelope.

\subsection{Model-form variations and components not included}

The nominal result uses the FiLM architecture and the $\lambda_{\rm ref}=1$
direct-$\FNP$ reference-distance objective.  Alternative penalties on
functional path length, deviations in $\log F_{\rm NP}$, curvature, and the
large-$b_T$ tail were examined separately.
None reduced the initialization dependence while preserving the fit quality
sufficiently to be adopted in the nominal analysis.  The quoted uncertainty
therefore remains conditional on the chosen model family, optimization
objective, perturbation family, and common empirical reference.

The combined envelope also does not include:
\begin{itemize}[leftmargin=1.3em]
  \item perturbative scale and profile variations;
  \item PDF-through-refit shifts of $\FNP$ or a final combination of NNPDF, CT,
  and MSHT high-$x$ systematics;
  \item nuclear-model variations beyond the fixed-target prescription;
  \item flavor-dependent $\FNP$ or alternative neural architectures;
  \item uncertainty from a complete experimental covariance model;
  \item fiducial-acceptance variations for the LHCb input; or
  \item a high-$q_T$ fixed-order-tail uncertainty outside the strict
  low-$q_T/Q$ fit domain.
\end{itemize}
It is therefore an empirical uncertainty envelope, not a complete global
QCD-theory, experimental, and model-form uncertainty.

\section{Results}
\label{sec:results}

\subsection{$b_T$-space TMDPDFs}

The fitted $\FNP$ is positive and monotone for all tabulated curves.  The full TMDPDFs exhibit a modest perturbative/OPE enhancement at small $\bt$, followed by nonperturbative damping at larger $\bt$.  The location and width of the broad maximum depend on $x$ and flavor through the perturbative PDF/OPE factor, even though $\FNP$ itself is flavor independent.

\begin{figure}[t]
\centering
\paperfigure{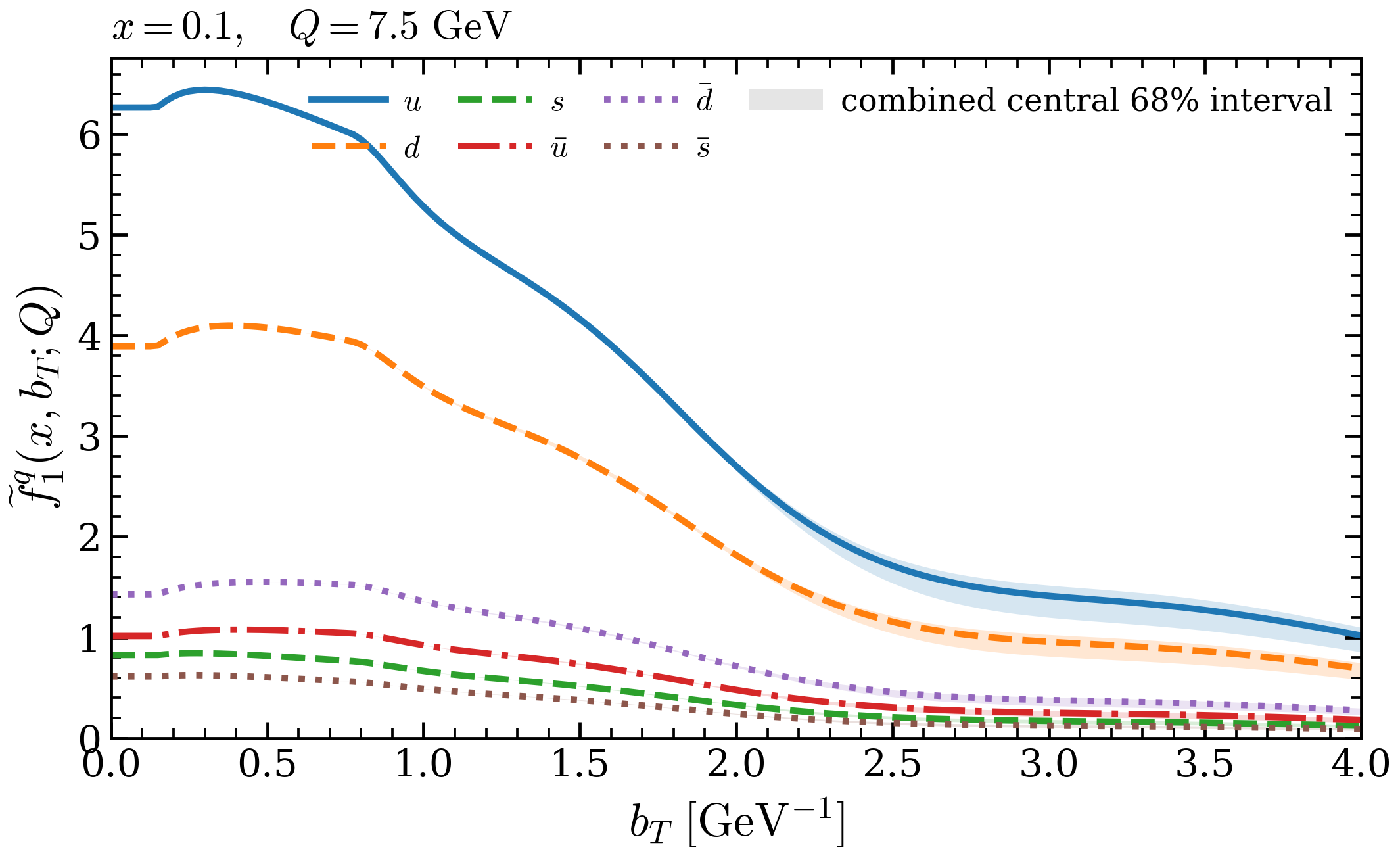}{1.0}
\caption{Representative impact-parameter-space TMDPDFs at $x=0.1$ and
$Q=7.5~\GeV$ for $u,d,s,\bar u,\bar d$, and $\bar s$.  Curves show the
medians of the combined ensemble, and shaded regions show its empirical
$q_{16}$--$q_{84}$ envelope formed from 96 converged, independently initialized fits and 50 experimental-residual fields.  The envelope is not a calibrated
one-standard-deviation interval.  The nonperturbative factor is shared by all
light flavors; flavor dependence arises from the collinear PDFs, perturbative
evolution, and OPE matching.  The single-TMD quantities exclude the hard factor.}
\label{fig:bspace}
\end{figure}

Figure~\ref{fig:bspace} shows the light-flavor $b_T$-space result at $x=0.1$ and $Q=7.5~\GeV$ for the nominal fit.  The curves are medians of the $4{,}800$ combined members, and the narrow shaded regions are the empirical $q_{16}$--$q_{84}$ combined envelopes.  The separate collinear-PDF uncertainty estimate is not included in this figure.

\begin{figure*}[t]
\centering
\paperfigure{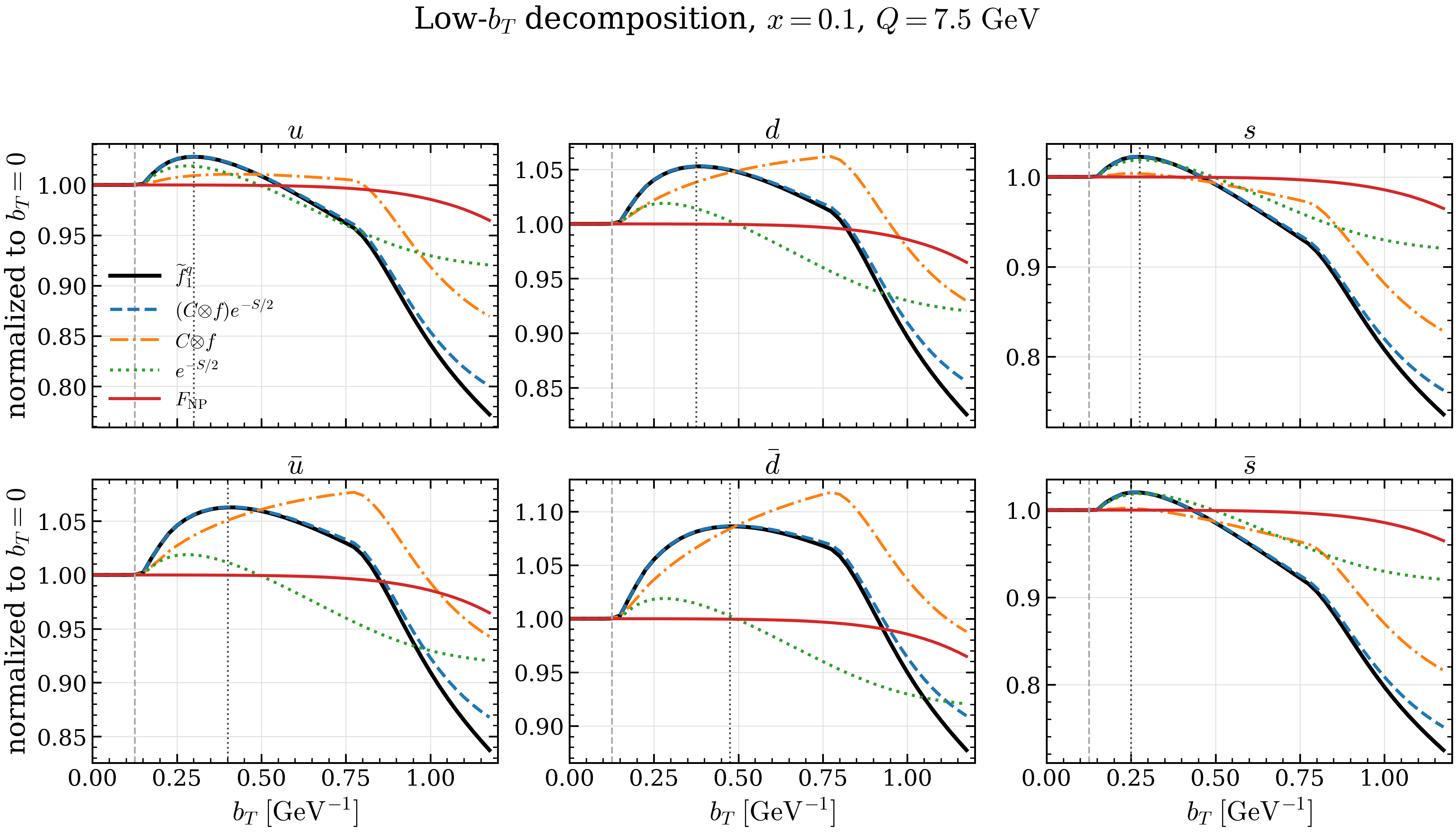}{1.0}
\caption{Decomposition of the low-$b_T$ hook and mild enhancement in the
nominal $q_T/Q\le0.20$ $b_T$-space TMDs.  The final TMD
$\widetilde f$ follows the perturbative/OPE/evolution components, while the
fitted $F_{\rm NP}$ remains monotone and close to unity in this region.  The
feature is therefore produced by the small-$b_T$ matching and profile scales,
not by an oscillatory nonperturbative factor.}
\label{fig:lowb-decomposition}
\end{figure*}

The small hook near $b_T \simeq 0.15~\mathrm{GeV}^{-1}$ and the mild enhancement over $b_T \simeq 0.25\text{--}0.48~\mathrm{GeV}^{-1}$ do not arise from an oscillatory structure learned by the nonperturbative factor. At the representative point $x=0.1$ and $Q=7.5~\mathrm{GeV}$, the central-grid result for $F_{\mathrm{NP}}$ is monotonic and remains close to unity throughout the region containing these features. Their origin is instead associated with the perturbative profile, operator-product-expansion matching, and evolution factors. Near $b_T=0$, $b_{\mathrm{pert}}$ is held close to the lower limit of the small-$b_T$ profile, while $\mu_b$ remains capped at $Q$ up to $b_T \simeq 0.125~\mathrm{GeV}^{-1}$. Above this transition, the OPE boundary condition and the single-leg Sudakov factor produce a modest enhancement before the damping at larger $b_T$ becomes dominant. This interpretation is checked directly in Fig.~\ref{fig:lowb-decomposition}, where the full TMD is decomposed into its perturbative/OPE/evolution and fitted nonperturbative components.  The observed structure should therefore be identified as a small-$b_T$ matching and profile feature, rather than interpreted as direct nonperturbative structure.

The flavor dependence of the displayed TMDPDFs should not be interpreted as
arising from independent flavor-dependent nonperturbative extractions.  The
model uses a single shared $F_{\rm NP}(x,b_T)$ for all light flavors.  The
differences among the $u$, $d$, $s$, $\bar u$, $\bar d$, and $\bar s$ curves
therefore arise from their respective collinear PDFs, OPE matching, and
perturbative evolution, each multiplied by the same fitted nonperturbative
factor.

% The plotted strange and antistrange curves should not be interpreted as an independent flavor-dependent strange nonperturbative extraction.  The model uses one shared $\FNP$ for all light flavors.  The strange curves are the strange collinear PDF/OPE/evolution legs multiplied by the same fitted nonperturbative factor.

\subsection{Cross-section description}

\begin{figure*}[t]
\centering
\paperfigure{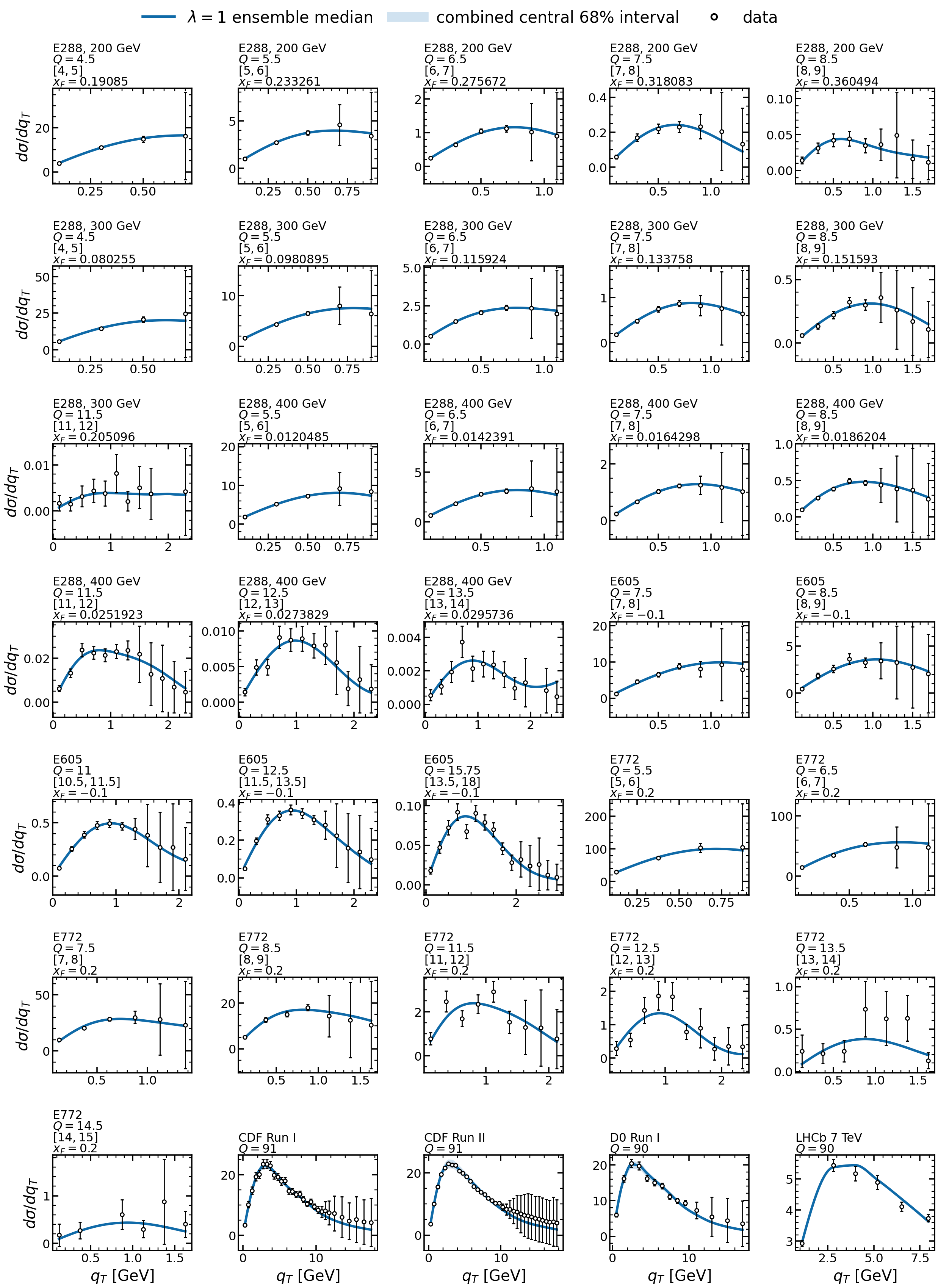}{0.86}
\caption{Comparison of the nominal $q_T/Q\le0.20$ data set with the
fitted $b_T$-space model.  Fixed-target curves are dense-grid evaluations of the
$W(b_T)$ kernels multiplied by the fitted nonperturbative factor; collider
curves use the published bin definitions, with interpolation only as a visual
guide.  The curves are combined-ensemble medians and the shading is the empirical
$q_{16}$--$q_{84}$ envelope from 96 converged, independently initialized fits and
50 experimental-residual fields.  Collinear-PDF uncertainty is not included.}
\label{fig:cross-sections}
\end{figure*}

Figure~\ref{fig:cross-sections} compares all nominal data in Table~\ref{tab:data} with smooth $q_T$ evaluations of the same fitted $b_T$-space model.  The shaded regions are the empirical $q_{16}$--$q_{84}$ combined envelopes over 96 converged, independently initialized fits and 50 experimental-residual fields.  They exclude collinear-PDF uncertainty, which is shown separately because it can be much larger in selected high-$x$ fixed-target panels.

The canvas is meant to test the observable-level consequences of the extracted $b_T$-space TMD, not only the appearance of the TMD itself.  The fixed-target panels show that the nominal fit gives a smooth representation of the measured $q_T$ spectra across the E288, E605, and E772 mass and beam-energy bins.  The collider panels provide a separate check that the Tevatron absolute spectra and the forward LHCb subset are described in their published bin conventions without requiring an additional shape deformation of the nonperturbative factor.  The remaining visible differences are therefore interpreted as ordinary residuals under the specified uncertainty model, not as evidence of a broken unit conversion or a non-smooth cross-section interpolation.

\begin{figure*}[t]
\centering
\paperfigure{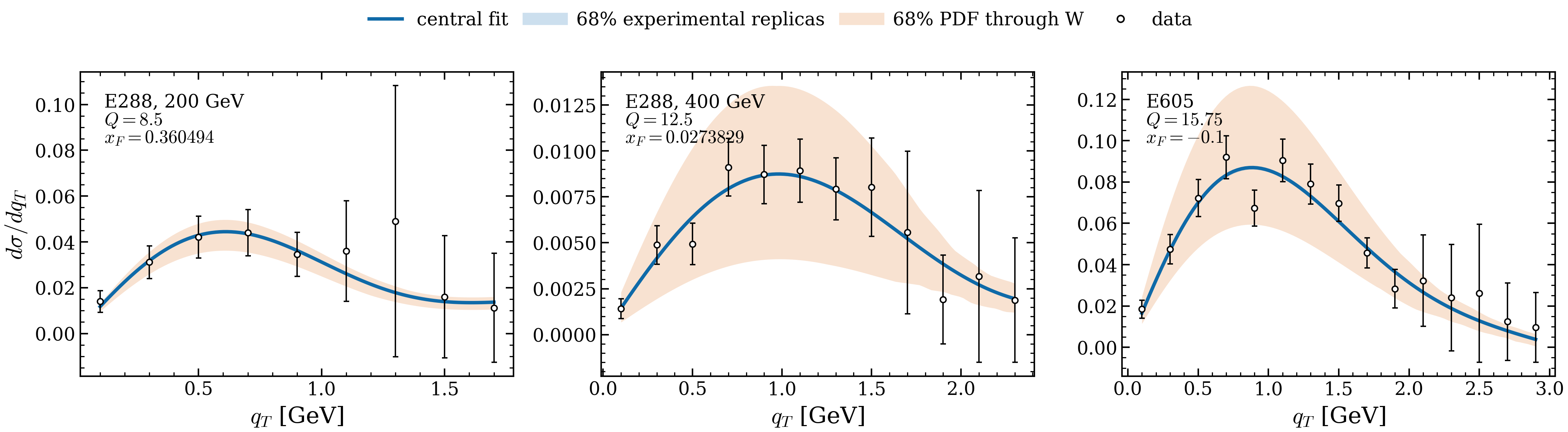}{1.0}
\caption{Selected cross-section panels comparing the experimental
$q_{16}$--$q_{84}$ envelope, evaluated with the initialization component fixed,
with the collinear-PDF variation obtained by recomputing $W(b_T)$ at fixed
central $F_{\rm NP}$.  In high-$x$ fixed-target kinematics the PDF luminosity
uncertainty can be much larger than the experimental shape component.}
\label{fig:pdf-example}
\end{figure*}

Figure~\ref{fig:pdf-example} isolates a different uncertainty question from the one addressed by Fig.~\ref{fig:cross-sections}.  In these selected panels the experimental-residual envelope, with the initialization component held fixed, is compared with a full-$W(b_T)$ PDF-member recalculation at fixed central $F_{\rm NP}$.  The much larger PDF band in high-$x$ fixed-target kinematics shows that the collinear luminosity uncertainty can dominate the observable uncertainty even when the fitted nonperturbative factor is tightly constrained.  This is why the PDF uncertainty estimate is not folded into the combined envelope.

\subsection{Regularized $k_T$-space representation}
\label{sec:kspace}
\begin{figure}[t]
\centering
\paperfigure{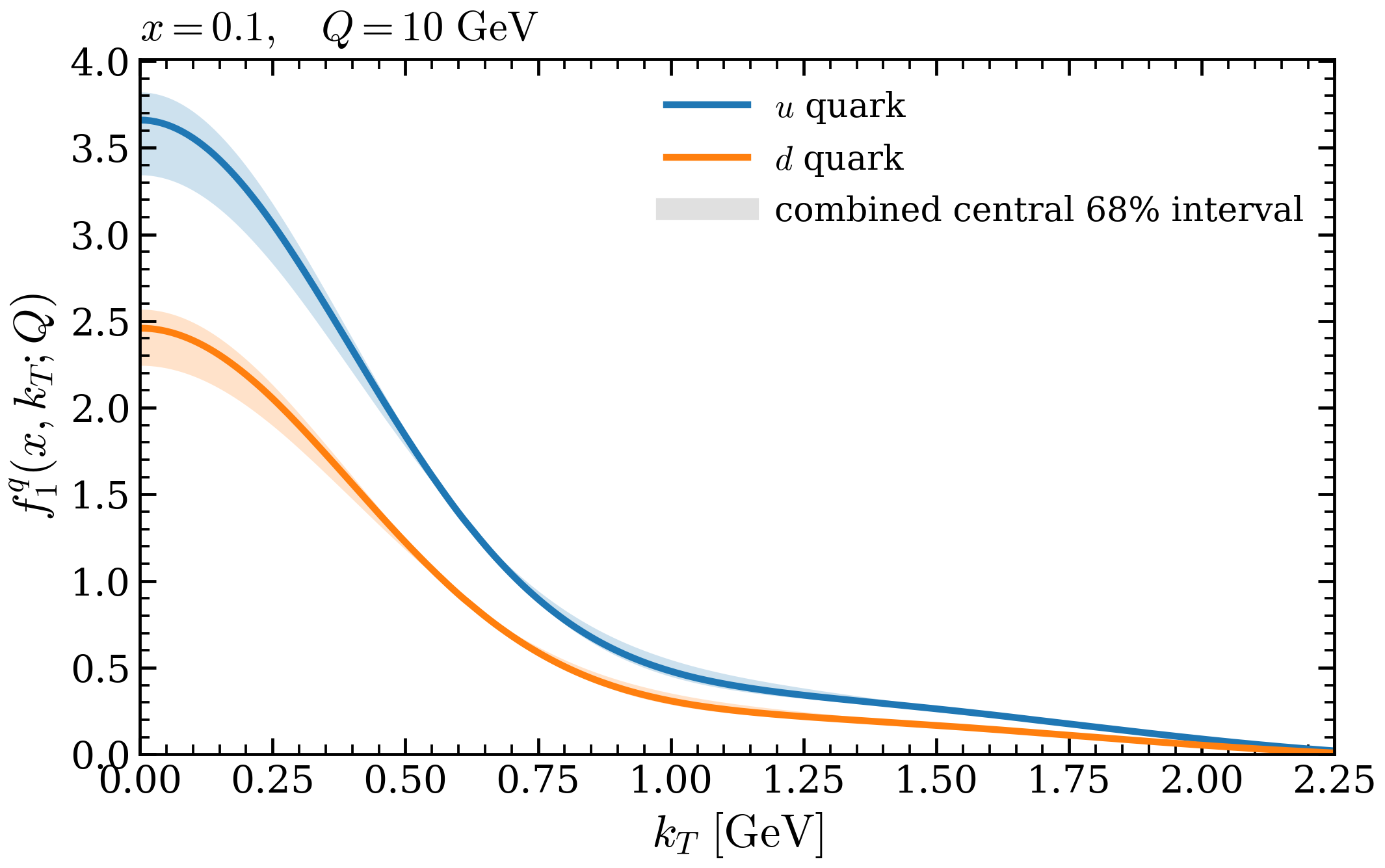}{1.0}
\caption{Regularized momentum-space representation of the fitted $u$- and
$d$-quark TMDPDFs at $x=0.1$ and $Q=10~\GeV$.  The curves are finite-$b_T$
Hankel transforms of the fitted $b_T$-space TMDs.  The shading is the empirical
$q_{16}$--$q_{84}$ envelope of the combined ensemble and is not interpreted as a
calibrated one-standard-deviation interval.}
\label{fig:kspace-fixedx}
\end{figure}

The fixed-$x$ momentum-space representation in
Fig.~\ref{fig:kspace-fixedx} gives a one-dimensional view of the regularized
Hankel transform for the two valence-dominated flavors at $x=0.1$ and
$Q=10~\GeV$.  The plotted range is restricted to the interval in which the
transform is stable and the median is positive for both flavors.  The $u$
distribution is larger and somewhat broader than the $d$ distribution,
reflecting the same collinear-PDF and OPE flavor hierarchy visible in
$b_T$ space.  The narrow combined envelope is consistent with the $b_T$-space
propagation: common normalization fluctuations are largely absorbed by
nuisance parameters, while the initialization and experimental variations of
$F_{\rm NP}$ remain small in this region.

\begin{figure*}[t]
\centering
\begin{minipage}[t]{0.485\textwidth}
\centering
\paperfigure{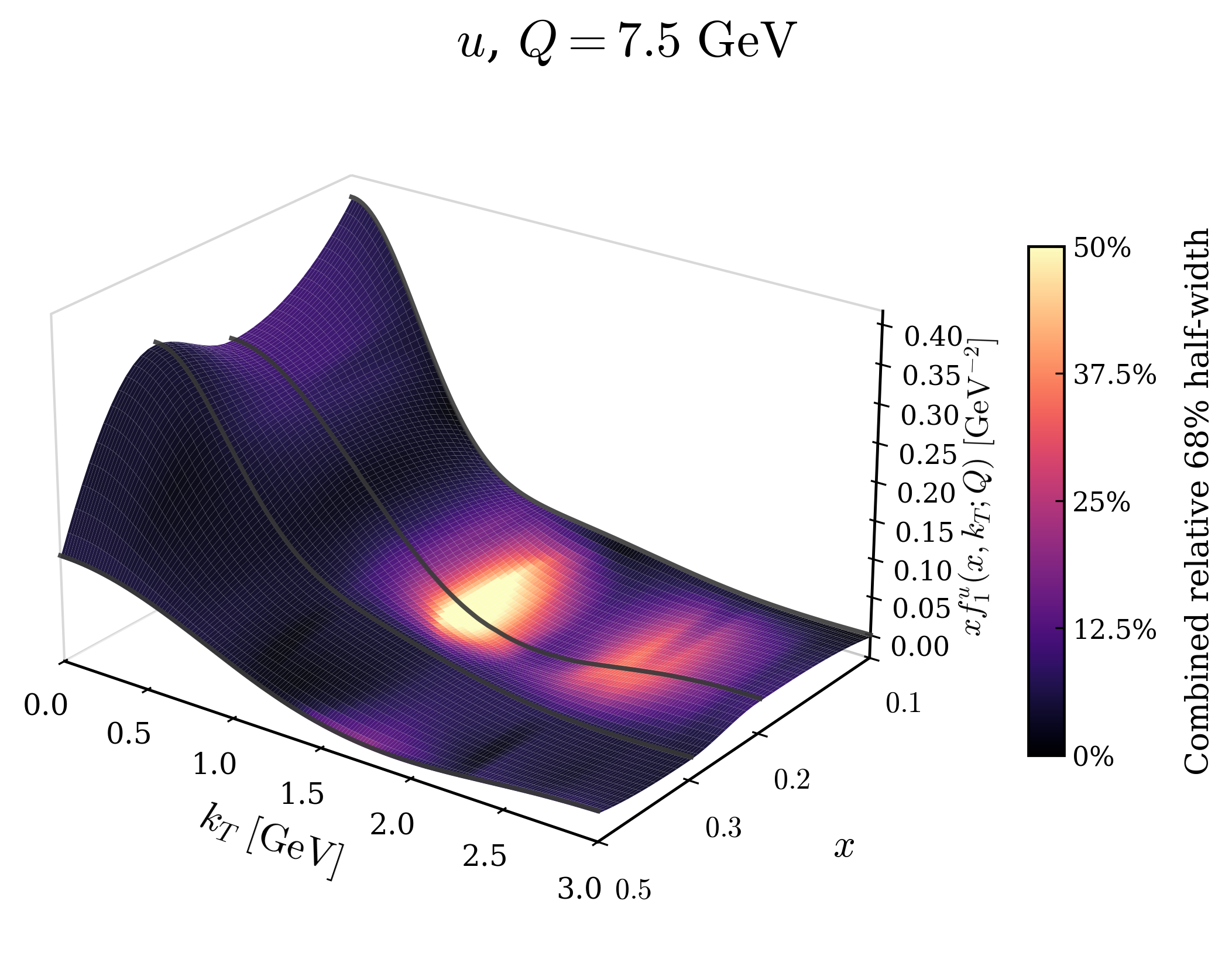}{0.96}
\caption{Regularized momentum-space companion TMDPDF
$x f_{1}^{u}(x,k_T;Q)$ at $Q=7.5~\GeV$.  The color scale is the pointwise
relative half-width of the empirical $q_{16}$--$q_{84}$ combined envelope from
96 converged, independently initialized fits and 50 experimental-residual fields; it is not a calibrated
one-standard-deviation band.}
\label{fig:kspace-surface-u}
\end{minipage}\hfill
\begin{minipage}[t]{0.485\textwidth}
\centering
\paperfigure{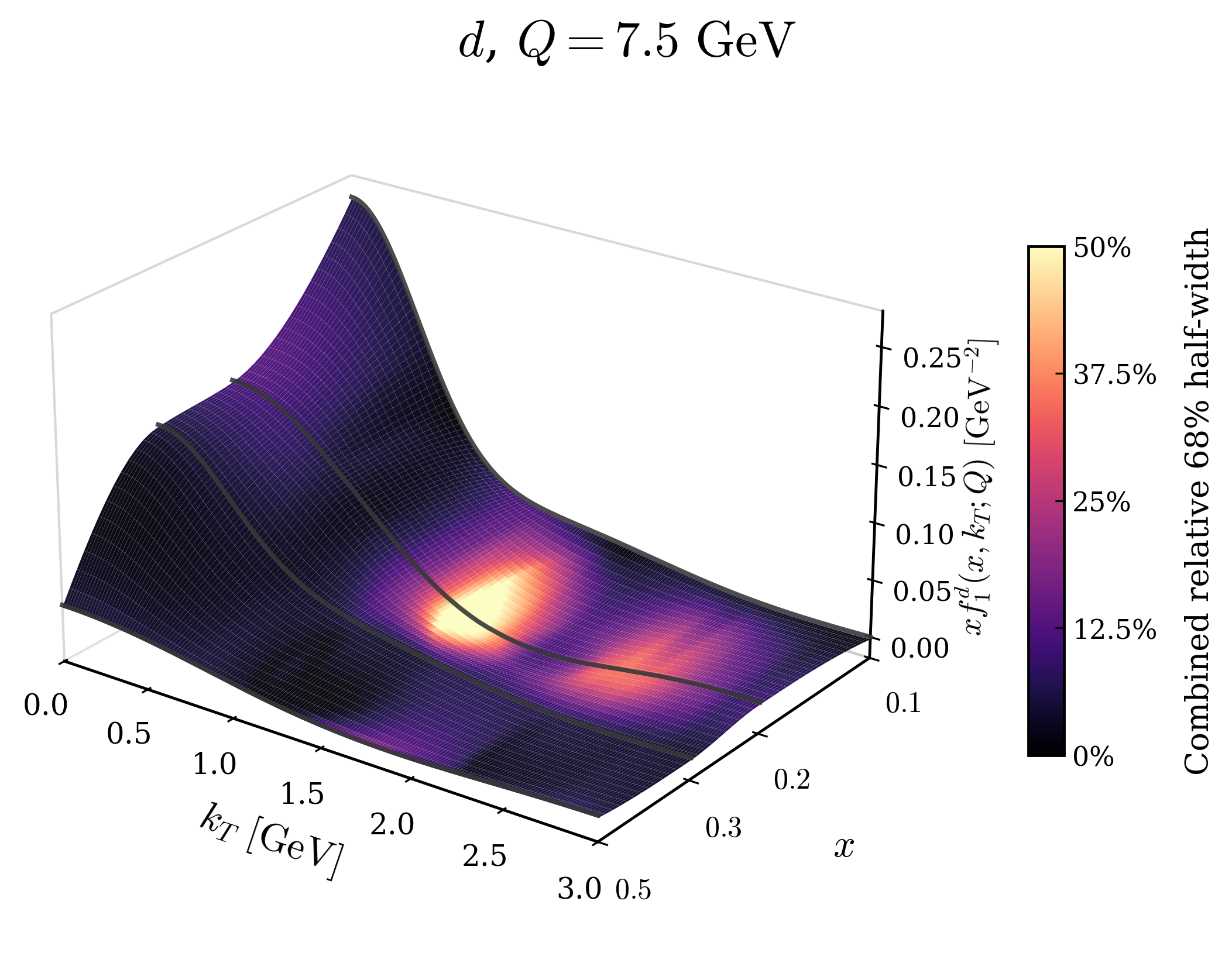}{0.96}
\caption{Regularized momentum-space companion TMDPDF
$x f_{1}^{d}(x,k_T;Q)$ at $Q=7.5~\GeV$.  The color scale is the pointwise
relative half-width of the empirical $q_{16}$--$q_{84}$ combined envelope from
96 converged, independently initialized fits and 50 experimental-residual fields; it is not a calibrated
one-standard-deviation band.  This is the $d$-flavor analogue of
Fig.~\ref{fig:kspace-surface-u}, with the same finite-$b_T$ transform
prescription and interpretation.}
\label{fig:kspace-d}
\end{minipage}
\end{figure*}
Figures~\ref{fig:kspace-surface-u} and \ref{fig:kspace-d} show the same
regularized transform as two-dimensional surfaces in $(x,k_T)$ for the $u$ and
$d$ flavors.  The color scale is the pointwise relative half-width of the
combined $q_{16}$--$q_{84}$ envelope, normalized to the magnitude of the
ensemble median with the specified numerical floor.  The surfaces show the
expected migration of support with $x$ and a rapid falloff at large $k_T$, but
they are finite-$b_T$ transforms of the fitted $b_T$-space distributions and
should not be interpreted as fixed-order high-$k_T$ predictions.

Large relative uncertainties in some high-$x$ sea-quark curves arise because their absolute TMDs are very small.  For the principal valence and moderate-$x$ curves the active-region relative bands are much smaller.  We therefore present the $k_T$-space representation as a regularized companion to the primary extraction in $b_T$ space.

The primary result is the $b_T$-space extraction.  Equation~\eqref{eq:hankel}
and the continuation described in Appendix~\ref{app:hankel} provide a
regularized momentum-space representation for comparison, but the physics
conclusions are based on the fitted $b_T$-space TMDs and their cross-section
predictions.

\subsection{Relation to direct momentum-space learning}

The present analysis and Ref.~\cite{FernandoKeller2025} share a
physics-informed neural-network strategy but assign different physics to the
learned function.  The direct momentum-space analysis learns an effective
transverse profile inside a differentiable convolution without an explicit
resummed Sudakov kernel.  Here the perturbative $W$ term, matching coefficients,
and evolution are fixed, and the network learns only $F_{\rm NP}$.  A direct
Fourier--Bessel comparison of the learned functions would therefore mix
different factorization and representation conventions.

A quantitative comparison requires the two analyses to use the same data and
kinematic support, perturbative accuracy, factorization and renormalization
scheme, definition of the learned object, and uncertainty construction.  Once
these ingredients are matched, one can vary only the momentum- or
impact-parameter-space representation and compare the resulting cross sections
and nonperturbative functions.  Without such matching, the meaningful comparison
is phenomenological: both methods can be tested against the same low-$q_T$
spectra and examined for consistent $x$ and scale dependence, but their learned
functions should not be identified with one another.

\section{Systematic studies and finite-$Y$ analyses}
\label{sec:systematics}

The main $b_T$-space and cross-section figures show the empirical
$q_{16}$--$q_{84}$ envelope formed from 96 converged, independently initialized fits and 50 experimental-residual fields.  The initialization,
experimental, and collinear-PDF components are also examined separately.  This
section establishes the kinematic range of the nominal fit, examines the
fixed-order and finite-$Y$ contributions near its upper boundary, and reports a
separate $W+Y$ candidate used to test model identifiability without redefining
the nominal extraction.

\subsection{Factorization-validity uncertainty}
\label{sec:factorization-validity}

The leading-power TMD factorization formula is an asymptotic
approximation in the region $q_T \ll Q$.  Contributions omitted from the
leading-power $W$-term description are therefore suppressed by powers of
the expansion parameter
\begin{equation}
  r_i \equiv \frac{q_{T,i}}{Q_i},
\end{equation}
together with possible hadronic power corrections involving
$\Lambda_{\mathrm{QCD}}/Q_i$.  This power-counting statement follows the
standard treatment of the remainder of TMD factorization
\cite{CollinsBook}.  It does not, by itself, determine the coefficient
or the detailed interpolation of the missing-power contribution at finite
$r_i$.

To test the stability of the extraction when data beyond the strict
$r_i\leq r_0$ region are included, we introduce a phenomenological
factorization-validity uncertainty.  For each included point, the published
diagonal experimental uncertainty is replaced in the fit by the effective
uncertainty
\begin{equation}
  \sigma_{i,\mathrm{eff}}^2
  =
  \sigma_{i,\mathrm{exp}}^2
  +
  \sigma_{i,\mathrm{fact}}^2,
  \label{eq:sigma-eff-factorization}
\end{equation}
where
\begin{equation}
  \sigma_{i,\mathrm{fact}}
  =
  |D_i|\,\delta_{\mathrm{fact}}(r_i).
  \label{eq:sigma-factorization}
\end{equation}
Here $D_i$ is the published central value of the measured cross section.
Thus, $\sigma_{i,\mathrm{fact}}$ is a fixed, pointwise theory uncertainty
constructed before fitting.  It is not an experimental uncertainty assigned
by the corresponding collaboration, and it is not recomputed from the
evolving theory prediction during minimization.  The same fixed effective uncertainties are used to generate the pseudo-data from which the experimental-residual fields are obtained.

We parameterize the relative factorization-validity uncertainty as
\begin{equation}
  \delta_{\mathrm{fact}}(r)
  =
  \lambda_{\rm fact}\,
  S\!\left(\frac{r-r_0}{w}\right)
  \left(\frac{r}{r_0}\right)^p,
  \label{eq:factorization-envelope}
\end{equation}
with the cubic smooth-step function
\begin{equation}
  S(z)=
  \begin{cases}
    0,             & z\leq 0,\\
    3z^2-2z^3,     & 0<z<1,\\
    1,             & z\geq 1.
  \end{cases}
  \label{eq:factorization-smoothstep}
\end{equation}
The threshold $r_0$ defines the strict leading-power core, while the width
$w$ avoids an abrupt change in statistical weight at its boundary.  The
power $p$ controls the assumed growth of the neglected contribution, and
$\lambda_{\rm fact}$ sets the overall size of the envelope.  Equation
\eqref{eq:factorization-envelope} is therefore our phenomenological
implementation of the power-counting expectation; it is not a unique formula
prescribed by the Collins formalism.  We treat the resulting contribution as
diagonal and independent between points.  No bin-to-bin or dataset-to-dataset
correlation model for the missing-power terms is introduced in the present
analysis.

The kinematic study enlarges the strict low-$q_T/Q$ sample in stages.  The strict fit uses only points satisfying $q_T/Q\leq 0.10$.
We first kept the fixed-target and LHCb selections at the strict cut while
extending the Tevatron collider points to $q_T/Q\leq 0.20$.  We then added
fixed-target points in stages up to $q_T/Q\leq 0.15$, $0.20$, and $0.25$.
For the staged study shown here we used
\begin{equation}
  r_0=0.10,
  \qquad
  w=0.05,
  \qquad
  \lambda_{\rm fact}=0.50,
  \qquad
  p=2.
  \label{eq:factorization-envelope-parameters}
\end{equation}
The value $r_0=0.10$ defines the boundary of the strict sample, and
$w=0.05$ provides a smooth interpolation into the enlarged-uncertainty region.
The choice $p=2$ implements quadratic growth.  The coefficient
$\lambda_{\rm fact}=0.50$ is fixed by requiring stability of the central
$b_T$-space shape and of the 50-member experimental-residual ensemble over the
parameter scan; it is not fitted to the cross-section data.  This coefficient
controls the factorization-validity uncertainty and is distinct from the
reference-distance weight $\lambda_{\rm ref}$ in
Eq.~\eqref{eq:objective-decomposition}.

For orientation, these parameters give $\delta_{\mathrm{fact}}(0.15)=1.125$,
$\delta_{\mathrm{fact}}(0.20)=2.000$, and
$\delta_{\mathrm{fact}}(0.25)=3.125$.
The prescription consequently downweights, rather than discards, points as
they move away from the strict leading-power core.  It should not be read as
a probabilistic derivation of the omitted contribution: it is a conservative
model-validity envelope whose adequacy is judged through the stability tests
described below.

The purpose of the scan was not to select the fit with the smallest
$\chi^2$.  The selection criterion is stability of the extracted $b_T$-space TMD from one
kinematic stage to the next, together with convergence of the
experimental-residual and initialization ensembles.  The strict sample contains
180 fitted points.  The Tevatron extension contains 216 points and remains
stable.  Adding all fixed-target points through $q_T/Q\leq0.15$ gives 278
points, and extending the same set through $q_T/Q\leq0.20$ gives 329 points.
These stages satisfy the $b_T$-space stability criteria and the 50-member
experimental-residual convergence tests.  The $q_T/Q\leq0.25$ stage increases
the fitted sample to 380 points and also satisfies the experimental-residual
test, but it crosses
the predefined applicability threshold
$\sigma_{\mathrm{fact}}/|D|=3$.

We therefore use the $q_T/Q\leq0.20$ fixed-target extension in the nominal analysis.  The $q_T/Q\leq0.25$ extension is shown only to assess the onset of sensitivity to higher-power or large-$q_T$ effects and is not used in the nominal result.
This distinction is important because the additional theory uncertainty can
reduce the apparent contribution to $\chi^2$ from marginal kinematics.  The
physics criterion is whether the extracted TMD remains stable under
controlled data enlargement, rather than whether the enlarged uncertainty
merely improves the goodness of fit.

The staged progression is summarized in Table~\ref{tab:factorization-progression}.
The tabulated envelope value is evaluated from
Eq.~\eqref{eq:factorization-envelope} at the largest sector-specific cut in each
stage.  The table displays the growth of the data set, the factorization-validity
uncertainty at each stage, and the final kinematic selection.

\begin{table*}[t]
\caption{Staged factorization-validity study.  The strict sample uses
$q_T/Q\leq0.10$ in every sector.  The fourth column gives
$\delta_{\rm fact}(r_{\max})=\sigma_{\rm fact}/|D|$ at the largest
sector-specific cut, using Eqs.~\eqref{eq:factorization-envelope} and
\eqref{eq:factorization-envelope-parameters}.  The nominal analysis uses the
329-point stage.  The 380-point extension exceeds the predefined
$\delta_{\rm fact}=3$ applicability threshold and is shown only as a systematic
comparison.  FT denotes the fixed-target data sets.}
\label{tab:factorization-progression}
\centering
\scriptsize
\begin{tabular}{lllll}
\toprule
Stage & \shortstack[l]{Sector-dependent\\selection} & $N$ & $\delta_{\rm fact}(r_{\max})$ & Status \\
\midrule
Strict sample
& All sectors: $r\leq0.10$
& 180 & 0 & Strict sample \\
Tevatron extension
& Tevatron: $r\leq0.20$; FT/LHCb: $r\leq0.10$
& 216 & 2.000 & Stability tests satisfied \\
Fixed-target extension
& Tevatron: $r\leq0.20$; FT: $r\leq0.15$; LHCb: $r\leq0.10$
& 278 & 2.000 & Stability tests satisfied \\
Nominal sample
& Tevatron/FT: $r\leq0.20$; LHCb: strict subset
& 329 & 2.000 & Included \\
Extended sample
& Tevatron: $r\leq0.20$; FT: $r\leq0.25$; LHCb: strict subset
& 380 & 3.125 & Not included \\
\bottomrule
\end{tabular}
\end{table*}

\subsection{High-$q_T$ fixed-order and $Y$ benchmarking}

The nominal fit does not include the higher-$q_T$ collider points.  The finite-$Y$ and fixed-order contributions are compared with DYTurbo and MCFM over representative Tevatron and LHC kinematics, including the same mass, rapidity, bin, and fiducial definitions used in the experimental tables.  We require
\begin{equation}
\frac{\left|\sigma_{\rm DYTurbo}-\sigma_{\rm MCFM}\right|}
{[\left|\sigma_{\rm DYTurbo}\right|+\left|\sigma_{\rm MCFM}\right|]/2}
\le 0.05 .
\end{equation}
The benchmark contains representative points from CDF Run I, CDF Run II, D0 Run I, and LHCb 7 TeV.  Fifteen DYTurbo/MCFM benchmark pairs were tested, including thirteen points with $q_T/Q>0.10$; all benchmarked pairs satisfy this criterion.  For LHCb, the MCFM fiducial prediction is scaled by $1/2$ before comparison because the implemented MCFM lepton cuts are symmetric in $|\eta|$ and therefore include both positive- and negative-rapidity hemispheres, while the LHCb measurement uses only the positive-rapidity hemisphere.

Table~\ref{tab:finite-tail-benchmark} summarizes this external-code comparison.
This comparison tests the normalization, binning, and fiducial conventions.  Agreement within 5\% shows that the two independent fixed-order implementations are consistent at the selected points, but it does not by itself justify including every higher-$q_T$ collider measurement in the nominal fit.

\begin{table}[t]
\caption{Representative DYTurbo/MCFM comparison.  The calculation uses
the same normalization, binning, and fiducial conventions as the experimental
data at the selected points.  The comparison does not imply that all
higher-$q_T$ measurements are included in the nominal fit.}
\label{tab:finite-tail-benchmark}
\centering
\scriptsize
\setlength{\tabcolsep}{3pt}
\begin{tabular}{@{}ll@{}}
\toprule
Comparison item & Result \\
\midrule
Independent calculations & DYTurbo and MCFM \\
Represented samples & \shortstack[l]{CDF Run I, CDF Run II, D0 Run I,\\LHCb 7 TeV} \\
\addlinespace[1.5pt]
Comparison points & 15 total \\
Points with $q_T/Q>0.10$ & 13 \\
\addlinespace[1.5pt]
Agreement criterion & Symmetric relative difference $\leq5\%$ \\
Outcome & 15 of 15 pairs satisfy the criterion \\
\addlinespace[1.5pt]
LHCb convention & \shortstack[l]{MCFM fiducial result divided by 2 for one\\forward rapidity hemisphere} \\
\bottomrule
\end{tabular}
\end{table}

Measurements outside the representative comparison are not included in the nominal fit unless their observable definition and fixed-order treatment are established separately.  The following subsection uses 24 additional Tevatron points only for a finite-$Y$ robustness test; the four higher-$q_T$ LHCb points remain excluded.

\subsection{Finite-$Y$ transition at the Tevatron boundary}
\label{sec:finite-y-boundary}

The fixed-order comparisons motivate a localized test of the finite-$Y$ contribution near the upper boundary of the Tevatron coverage.  We start from the nominal 329-point solution of Sec.~\ref{sec:method}, for which the finite-$Y$ correction remains zero, and add 24 Tevatron points extending to $q_T/Q\simeq0.30$.  A smooth profile interpolates between the resummed and fixed-order predictions only for these additional points.  The four higher-$q_T$ LHCb measurements remain excluded.

For $r=q_T/Q$, the transition is implemented as
\begin{align}
Y_{\rm unitary}(r)
&=p(r)\left[\sigma_{\rm FO}^{\rm NLO}(r)-\sigma_W(r)\right],
\label{eq:unitary-finite-y}\\
\sigma_{\rm matched}(r)
&=\sigma_W(r)+Y_{\rm unitary}(r)\nonumber\\
&=\left[1-p(r)\right]\sigma_W(r)
  +p(r)\sigma_{\rm FO}^{\rm NLO}(r),
\label{eq:unitary-matched}
\end{align}
Here $\sigma_W$ and $\sigma_{\rm FO}^{\rm NLO}$ are evaluated in the same
bin-level observable convention, so $Y_{\rm unitary}$ denotes the observable-level
finite-$Y$ correction, including their common prefactors.  The function $p(r)$ is a smooth profile.  It is zero on the nominal 329-point sample and becomes
nonzero only for the additional points.  The central profile is varied by moving the
transition window earlier and later, providing an estimate of the sensitivity
to the location of the $W$-to-fixed-order interpolation.  The construction is
unitary in the stated sense: the prediction remains $W$ in the core and
approaches the NLO fixed-order result as the profile is turned on.  It is not a
claim that the transition is a complete all-orders matching construction.  When
$F_{\rm NP}$ is refitted, convergence is assessed from the norm of the objective
gradient with respect to the nonperturbative parameters, denoted
$\|\nabla_{\theta_{\rm NP}}\Phi\|$, with a threshold of $10^{-4}$.

\begin{table*}[t]
\caption{Effect of the finite-$Y$ transition on the 24 additional
Tevatron points.  The nominal 329-point sample is unchanged and has $Y=0$.  In
the first three entries $F_{\rm NP}$ is held fixed while the normalization
nuisances are refitted; the fourth entry allows $F_{\rm NP}$ to vary.  The listed
$\chi^2/N$ values use the finite-$Y$ comparison objective and are not directly
comparable to the nominal-fit data term.}
\label{tab:finite-y-comparison}
\centering
\scriptsize
\begin{tabular}{lcccll}
\toprule
Profile & $F_{\rm NP}$ treatment & \shortstack{Total\\$\chi^2/N$} & \shortstack{Additional points\\$\chi^2/N$} & Convergence & Status \\
\midrule
Early   & fixed    & 0.4566 & 1.027 & Satisfied     & Included \\
Central & fixed    & 0.4599 & 1.074 & Satisfied     & Included \\
Late    & fixed    & 0.4615 & 1.100 & Satisfied     & Included \\
\midrule
Central & refitted & 0.4558 & 1.042 & Not satisfied & Excluded \\
\bottomrule
\end{tabular}
\end{table*}

When $F_{\rm NP}$ is held fixed and only the normalization nuisances are
refitted, all three optimizations converge.  The largest change in the prediction
for the nominal 329-point sample is approximately $1.2\%$.  Allowing
$F_{\rm NP}$ to vary lowers the comparison objective slightly, but the gradient
norm remains $\|\nabla_{\theta_{\rm NP}}\Phi\|\simeq1.40$, far above the
$10^{-4}$ convergence criterion.  This refit is therefore not used to define the
finite-$Y$ robustness result, and the nominal 329-point solution remains
unchanged.

This comparison provides a controlled extension of the tested Tevatron
region to $q_T/Q\simeq0.30$.  It does not establish a globally matched high-$q_T$
prediction, redefine the nominal sample, or alter the nominal
factorization-validity uncertainty model.

\subsection{\texorpdfstring{$W+Y$}{W+Y} candidate and identifiability study}
\label{sec:wy-candidate}

The boundary test of Sec.~\ref{sec:finite-y-boundary} modifies the prediction
only for 24 additional Tevatron points while preserving the nominal 329-point
$W$-only fit.  We consider two distinct calculations: a directly evaluated
perturbative $W+Y$ grid for the Tevatron measurements and the response of the
flexible FiLM extraction when the specified finite-$Y$ input is supplied on the
nominal data set.  These results are kept separate because the first is a
perturbative cross-section benchmark, whereas the second is a
model-identifiability study.

\paragraph*{Direct Tevatron $W+Y$ benchmark.}
An independent DYTurbo evaluation produced a directly evaluated
$\NthreeLL+\mathrm{NNLO}$ $W+Y$ grid for 122 Tevatron bins: 41 from CDF Run~I,
61 from CDF Run~II, and 20 from D0 Run~I.  The calculation uses the convention
\begin{align}
W &= \sigma_{\rm RES},\nonumber\\
Y &= \sigma_{\rm FO}^{\rm NNLO}-\sigma_{\rm ASY}^{\rm NNLO},\nonumber\\
W+Y &= \sigma_{\rm RES}+\sigma_{\rm CT}+\sigma_{V+\mathrm{jet}},
\label{eq:direct-wy-convention}
\end{align}
where $\sigma_{\rm RES}$ is the resummed contribution,
$\sigma_{\rm ASY}^{\rm NNLO}$ is its NNLO asymptotic expansion,
$\sigma_{\rm CT}$ is the corresponding counterterm, and
$\sigma_{V+\mathrm{jet}}$ is the fixed-order vector-boson-plus-jet
contribution.  In the nonperturbative convention used for this isolated
DYTurbo calculation, the parameter is fixed at $g_1=1.017~\GeV^2$.

The resulting table is finite and positive in every bin.  The ratio of the
calculated $W+Y$ cross section to the measured value has median 1.003 and ranges
from 0.775 to 1.168.  The estimated numerical integration uncertainty averages
1.18\% and reaches 16.5\% in bins dominated by cancellations.  Reconstructing
$Y$ explicitly from $\sigma_{\rm FO}^{\rm NNLO}-\sigma_{\rm ASY}^{\rm NNLO}$
and comparing it with the counterterm-plus-jet representation gives a median
residual of $2.95\times10^{-4}~\mathrm{pb}/\GeV$ and a maximum residual
of $9.07\times10^{-3}~\mathrm{pb}/\GeV$.  These integration estimates diagnose
the perturbative grid and are not added as statistical uncertainties on the
extracted TMD.  The $\NthreeLL+\mathrm{NNLO}$ designation in this paragraph
applies to this isolated DYTurbo grid; it does not alter the accuracy statement
for the nominal neural extraction.

\paragraph*{Candidate coupled to the 329-point fit.}
To study the response of $F_{\rm NP}$, a separate candidate was constructed from
an external $W$ table with consistent data-set/bin matching and fiducial-factor
application, together with an assembled finite-$Y$ correction.
The assembled candidate input contains 321 nonzero finite-$Y$ entries and eight
entries with $Y=0$.  Six of the zero entries are the retained LHCb points, which
remain $W$-only diagnostics.  The remaining two are non-LHCb rows for which the
tabulated finite-$Y$ correction vanishes.  The available LHCb
full-minus-resummed subtraction is
not used because cancellation-driven integration errors give both signs and
relative subtraction uncertainties as large as 234\%.  Using those central
values as exact fit inputs would allow numerical integration noise to generate
spurious structure in $F_{\rm NP}$.  The candidate is therefore neither a
finite-$Y$ treatment of every nominal point nor a fully matched 353-point
$W+Y$ extraction.

The candidate retains the positive-rate monotone FiLM architecture of
Sec.~\ref{sec:method}, but changes the regularization relative to the nominal
fit.  Its objective is
\begin{equation}
\Phi_{W+Y}
=
\chi^2_{\rm data}
+
\chi^2_{\rm norm}
+
3\,\chi^2_{\rm ref}
+
\chi^2_{\rm tail}.
\label{eq:wy-candidate-objective}
\end{equation}
The subscript $W+Y$ identifies the exploratory fit in which the available
finite-$Y$ contributions are included in the cross-section prediction; it does
not imply that $Y$ is nonzero for every fitted row.  The direct-$F_{\rm NP}$
reference distance uses the same denominator floor as
Eq.~\eqref{eq:reference-residual}, but is evaluated over
$0.1\le b_T\le8.0~\GeV^{-1}$ with coefficient
$\lambda_{\rm ref}=3$.

For completeness, let $\mathcal X_{\rm tail}$ denote the candidate $x$ grid
used for the large-$b_T$ control and let
$\mathcal B_{\rm tail}$ be the candidate $b_T$ nodes satisfying
$b_T\ge6.0~\GeV^{-1}$.  With
$N_{\rm tail}=|\mathcal X_{\rm tail}|\,|\mathcal B_{\rm tail}|$, the
implemented tail contribution is the accepted-row-scaled mean squared excess
above the target value 0.05,
\begin{equation}
\begin{aligned}
\chi^2_{\rm tail}
&=
\frac{N_{\rm acc}}{N_{\rm tail}}
\sum_{x_j\in\mathcal X_{\rm tail}}
\sum_{b_a\in\mathcal B_{\rm tail}}\\[-2pt]
&\quad\times
\left[
\max\!\left(F_{\rm NP}(x_j,b_a)-0.05,0\right)
\right]^2.
\end{aligned}
\label{eq:wy-tail-control}
\end{equation}
The reference and tail terms are soft controls: they do not impose a boundary
value at $b_T=8~\GeV^{-1}$ and do not make the inverse problem mathematically
unique.

The calculation uses 96 independently initialized fits obtained from 1\%
parameter perturbations, with seeds 303--398, and 50 pseudo-data residual fits,
with seeds 1001--1050.  Each optimization has a 50,000-epoch ceiling.  The
combined ensemble is formed by adding the centered pseudo-data residuals in
$\log F_{\rm NP}$ to each independently initialized curve, yielding
$96\times50=4{,}800$ members.  It therefore contains 96 fitted initialization
solutions and 50 fitted residual fields, not 4,800 independent neural-network
optimizations.  The quoted $q_{16}$, $q_{50}$, and $q_{84}$ values are empirical
quantiles of this construction.

There is an important convergence qualification.  Every candidate run reached
the 50,000-epoch ceiling; 26 of the 96 independently initialized fits and 48 of
the 50 pseudo-data fits have their best stored epoch at that ceiling.  The
candidate therefore does not possess the same explicit block-to-block
$F_{\rm NP}$ convergence certificate as the nominal ensemble.  This is one
reason it is treated as an exploratory diagnostic.

\paragraph*{Effect on the extracted distributions.}
The recorded scalar optimization metric per nominal data point has median
0.4152 over the independently initialized fits, with range
0.3126--0.4447, compared with 0.4030 for the nominal
$\lambda_{\rm ref}=1$ fits.  This is not a controlled fit-quality comparison:
the candidate simultaneously changes the external $W$ table, introduces the
non-LHCb finite-$Y$ input, and extends the reference-distance interval.

Table~\ref{tab:wy-width-comparison} compares the active-region empirical
$q_{16}$--$q_{84}$ full widths.  The stronger full-range reference term does not
produce a uniformly smaller envelope.  It reduces the spread near
$b_T\simeq3$--$4~\GeV^{-1}$, which controls the very-low-$k_T$ Hankel endpoint,
but broadens both the data-sensitive $b_T<2~\GeV^{-1}$ region and the
intermediate-$k_T$ region.

\begin{table}[t]
\caption{Active-region full widths of the empirical central 68\% intervals,
defined by $q_{84}-q_{16}$, for the nominal $\lambda_{\rm ref}=1$ ensemble
and the exploratory $W+Y$ candidate with $\lambda_{\rm ref}=3$.}
\label{tab:wy-width-comparison}
\centering
\scriptsize
\setlength{\tabcolsep}{4pt}
\begin{tabular}{lcc}
\toprule
Quantity & Nominal & $W+Y$ candidate \\
\midrule
$b_T$ space, $u/d$ maximum & 24.0\% & 29.8\% \\
$b_T$ space, $u/d$ median  &  3.1\% & 11.7\% \\
$k_T$ space, $u$ maximum  & 21.3\% & 28.5\% \\
$k_T$ space, $u$ median   & 11.1\% & 13.9\% \\
$k_T$ space, $d$ maximum  & 22.5\% & 27.7\% \\
$k_T$ space, $d$ median   & 11.5\% & 14.3\% \\
\bottomrule
\end{tabular}
\end{table}

At representative points, the candidate full width is 8.3\% near
$b_T=1~\GeV^{-1}$, 27.6\% near $b_T=2~\GeV^{-1}$, and 10.5\% near
$b_T=4~\GeV^{-1}$.  Relative to the nominal median, the candidate
$F_{\rm NP}$ is lower by approximately 6--8\% over
$b_T=1$--$2~\GeV^{-1}$ and higher by approximately 8\% near
$b_T=4~\GeV^{-1}$.  The change is therefore a shift in the central shape as
well as in the uncertainty envelope.

\begin{figure*}[t]
\centering
\begin{minipage}[t]{0.485\textwidth}
\centering
\paperfigure{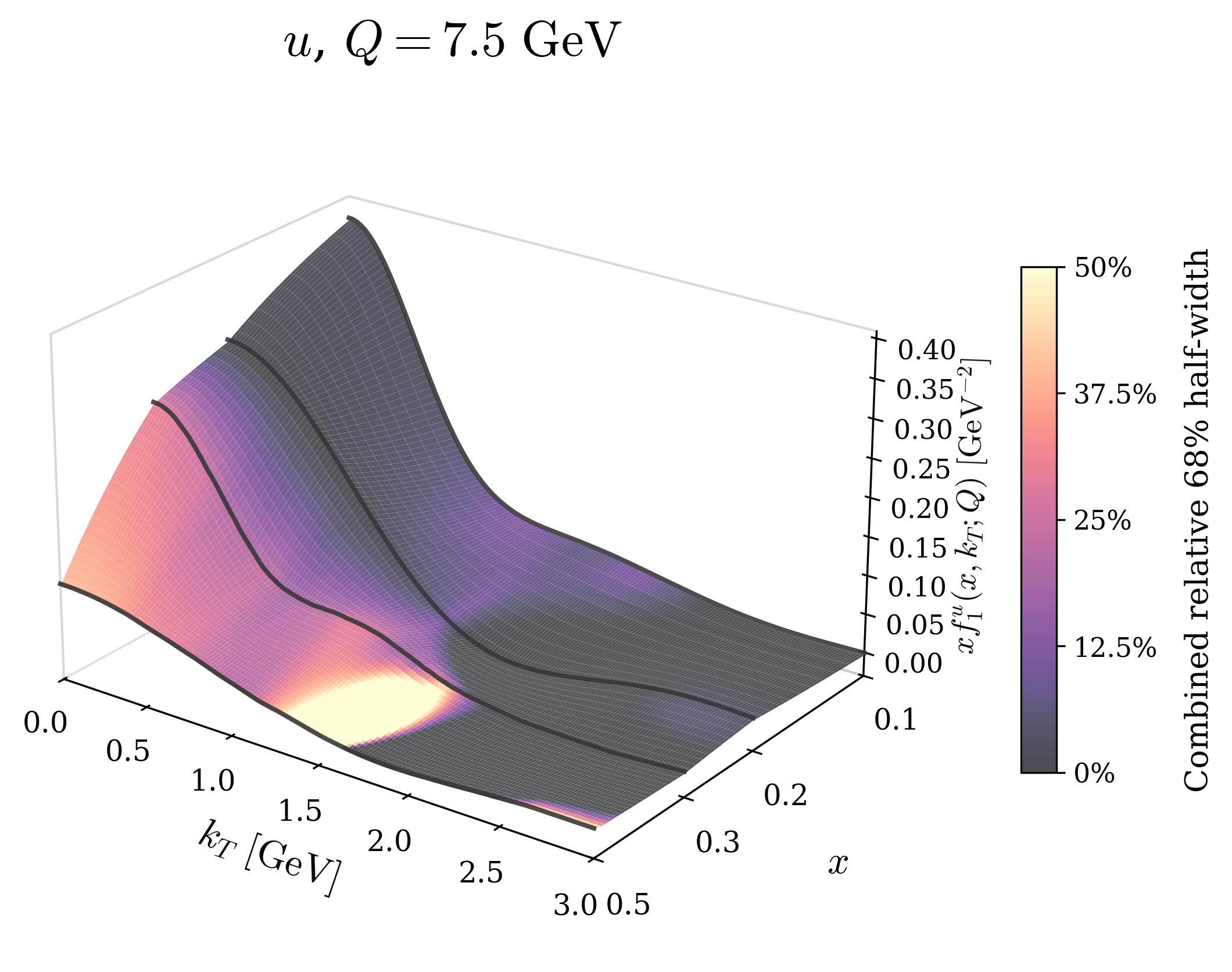}{0.98}
\caption{The $W+Y$ candidate for
$x f_1^u(x,k_T;Q)$ at $Q=7.5~\GeV$.  The surface height is the empirical
$q_{50}$ from 96 independently initialized fits crossed with 50 residual fields,
and the color is the relative half-width of its empirical central 68\% interval,
defined by $q_{16}$ and $q_{84}$.  The direct numerical $x$ nodes are 0.1, 0.2,
0.3, and 0.5; interpolation between them is for display only. }
\label{fig:wy_candidate_u_surface}
\end{minipage}\hfill
\begin{minipage}[t]{0.485\textwidth}
\centering
\paperfigure{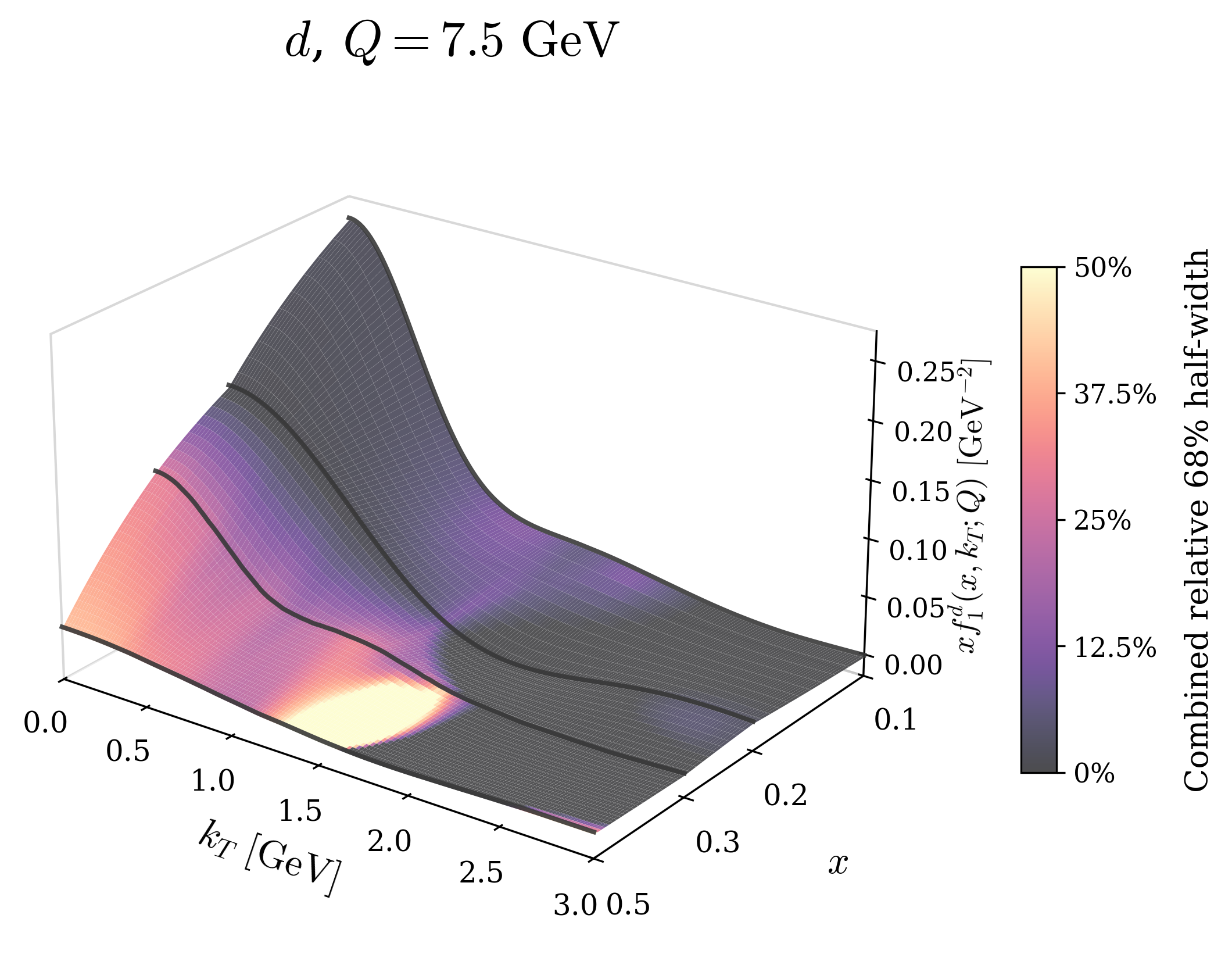}{0.98}
\caption{The $W+Y$ candidate for
$x f_1^d(x,k_T;Q)$ at $Q=7.5~\GeV$, constructed with the same specified
non-LHCb finite-$Y$ input and 96-by-50 ensemble prescription as
Fig.~\ref{fig:wy_candidate_u_surface}.  The surface height is the empirical
$q_{50}$ and the color is the relative half-width of the empirical central
68\% interval defined by $q_{16}$ and $q_{84}$.}
\label{fig:wy_candidate_d_surface}
\end{minipage}
\end{figure*}

Figures~\ref{fig:wy_candidate_u_surface} and
\ref{fig:wy_candidate_d_surface} show the corresponding regularized
momentum-space surfaces at $Q=7.5~\GeV$.  The direct numerical $x$ nodes are
$x=0.1$, 0.2, 0.3, and 0.5; the denser interpolation in $x$ is used only to
render a smooth surface.  The height is the empirical median, and the color is
the relative half-width of the empirical central 68\% interval, defined by the
$q_{16}$ and $q_{84}$ quantiles.  The label ``Combined relative 68\%
half-width'' retained in the color bar denotes this operational interval and
does not assign a calibrated Gaussian confidence interpretation.

The exploratory calculation therefore shows that stronger full-range reference
control relocates the nonuniqueness rather than eliminating it.  A causal test
of the finite-$Y$ contribution alone would require repeating the
$\lambda_{\rm ref}=1$ analysis with exactly the same $W$ and non-LHCb
finite-$Y$ inputs.  In the absence of that controlled comparison and a
convergence certificate for the candidate fits, the nominal $W$-only ensemble
remains the primary extraction.

\subsection{Pseudo-data and inverse-transform closure}
\label{sec:pseudodata-closure}

The experimental component of the combined ensemble is derived from 50
pseudo-data fits.  We therefore test both the pseudo-data generation and the
numerical Fourier--Bessel transform used for the regularized momentum-space
representation.  These tests address the statistical and numerical consistency
of the nominal $q_T/Q\le0.20$ analysis.

\paragraph*{Observable-level zero closure.}
The first test is a zero-level closure test in the exact data-point and bin convention
used by the nominal fit.  For each nominal data point \(i\), we replace the
measured target by the central fitted prediction,
\begin{equation}
D_i^{(0)} = T_i^{\rm cent},
\end{equation}
where \(T_i^{\rm cent}\) is the nominal fitted prediction after the same bin
convention, data-set normalization factor, power-counting-motivated factorization-validity
uncertainty, and theoretical convention used in the fit.  The closure residual is
\begin{equation}
r_i^{(0)}
=
\frac{D_i^{(0)}-T_i^{\rm cent}}{\sigma_{i,\rm eff}} .
\end{equation}
This test checks data-point ordering, unit conventions, the fitted bin-level observable,
the use of \(\sigma_{i,\rm eff}\), and the prediction column used downstream by
the pseudo-data generator.  For the 329 points in the nine data sets, the result is
\begin{equation}
\max_i |r_i^{(0)}| = 0,
\qquad
\frac{1}{N}\sum_i \left(r_i^{(0)}\right)^2 = 0 .
\end{equation}
Thus the observable-level pseudo-data table closes exactly when the input data
are generated from the fitted theory itself.  This is a necessary check because
a mismatch of a Jacobian, normalization convention, data-point ordering, or collider
bin definition would usually appear as a nonzero residual even before adding
statistical noise.

\paragraph*{Pseudo-data pull calibration.}
The second test verifies that pseudo-data fluctuations on the scale used in the
fit produce correctly calibrated pulls.  We generated \(10^4\) pseudo-data
tables from the central fitted prediction.  Since the nominal sample has
329 points, each ensemble contains \(3.29\times10^6\) point-level pulls.
Two toy prescriptions were tested.

The first prescription uses only the point-level effective uncertainty,
\begin{equation}
D_i^{(r)}
=
T_i^{\rm cent}
+
\sigma_{i,\rm eff}\,z_i^{(r)},
\qquad
z_i^{(r)}\sim{\cal N}(0,1),
\end{equation}
with pulls
\begin{equation}
p_i^{(r)}
=
\frac{D_i^{(r)}-T_i^{\rm cent}}{\sigma_{i,\rm eff}} .
\end{equation}
Here \(\sigma_{i,\rm eff}\) is the same effective point uncertainty used in the
nominal data table, including the power-counting-motivated factorization-validity component
where applicable.  This gives
\begin{align}
\langle p\rangle &= -3.0\times10^{-4},\\
\mathrm{std}(p) &= 1.0003,\\
\left\langle \chi^2/N\right\rangle &= 1.0006 .
\end{align}
The empirical fractions of pulls satisfying \(|p|\le1\) and \(|p|\le2\) are
\begin{equation}
0.6827
\quad\mathrm{and}\quad
0.9545 ,
\end{equation}
respectively, in agreement with the expected Gaussian coverages.

The second prescription adds a data-set-wide normalization fluctuation on top of
the point-level effective fluctuations,
\begin{equation}
D_i^{(r)}
=
\left(1+\delta_{d(i)} z_{d(i)}^{(r)}\right)T_i^{\rm cent}
+
\sigma_{i,\rm eff}\,z_i^{(r)} ,
\end{equation}
where \(d(i)\) labels the data set of point \(i\), \(\delta_d\) is the
normalization width used for that dataset, and \(z_d^{(r)}\) is a common
standard-normal draw for the data set.  For this calibration test the pull
denominator is the quadrature sum
\begin{equation}
\sigma_{i,\rm toy}
=
\left[
\sigma_{i,\rm eff}^2
+
\left(\delta_{d(i)}T_i^{\rm cent}\right)^2
\right]^{1/2}.
\end{equation}
The resulting pull distribution has
\begin{align}
\langle p\rangle &= 1.5\times10^{-3},\\
\mathrm{std}(p) &= 1.0011,\\
\left\langle \chi^2/N\right\rangle &= 1.0022 ,
\end{align}
with empirical one- and two-standard-deviation coverages
\begin{equation}
0.6822
\quad\mathrm{and}\quad
0.9542 .
\end{equation}
The toy-to-toy spread of \(\chi^2/N\) is larger in this second prescription
because points in the same data set receive a common-mode normalization shift.  This
is expected and is not a failure of the point-level uncertainty model.

\begin{table}[t]
\caption{Pseudo-data pull-closure summary for the nominal \(q_T/Q\le0.20\) fit.  Each column uses \(10^4\) pseudo-data toys, corresponding to
\(3.29\times10^6\) point-level pulls.}
\label{tab:pseudodata-closure}
\begin{ruledtabular}
\begin{tabular}{lcc}
Metric & pointwise only & normalization + pointwise\\
\hline
\(\langle p\rangle\) & \(-0.0003\) & \(0.0015\)\\
std.\((p)\) & \(1.0003\) & \(1.0011\)\\
\(\langle\chi^2/N\rangle\) & \(1.0006\) & \(1.0022\)\\
\(P(|p|\le1)\) & \(0.6827\) & \(0.6822\)\\
\(P(|p|\le2)\) & \(0.9545\) & \(0.9542\)\\
\end{tabular}
\end{ruledtabular}
\end{table}

These numbers have a simple statistical interpretation.  If the specified uncertainties are internally consistent with the pseudo-data
generator, the standardized pulls should be distributed as a unit normal up to
Monte Carlo fluctuations.  With \(3.29\times10^6\) pulls, the observed deviations
of the mean, width, and Gaussian coverages from their ideal values are at the
\(10^{-3}\) level.  The pseudo-data generation therefore propagates fluctuations with the intended
scale; the relatively small experimental component is not caused by
an accidentally suppressed pseudo-data fluctuation.

\paragraph*{Relation to the 50 experimental-residual fields.}
The pseudo-data pull tests establish the calibration of the stochastic input
to the experimental component.  They do not, by themselves, establish the
behavior of the neural-network refits.  For the 50 pseudo-data fits used to
construct the experimental-residual fields,
\begin{align}
\mathrm{median}(\overline\chi^{2}) &= 1.4521,\\
q_{0.95}(\overline\chi^{2}) &= 1.6028,\\
q_{0.95}(|p_{\rm norm}|) &= 2.9709,
\end{align}
and the active-region split-half comparisons give
\begin{align}
q_{0.95}(d_{\rm center}) &= 0.0132,\\
q_{0.95}(d_{\rm width}) &= 0.4512 .
\end{align}
These results show that the pseudo-data fluctuations have the intended
calibration and that the 50-member experimental-residual ensemble is stable
under the split-half comparison.  They do not assign a confidence level to the
combined 96-by-50 envelope.  The test is conditional
on the nominal model class and training protocol; it does not include
architecture, scale-profile, nuclear-model, covariance, or PDF-through-refit
variations.

\paragraph*{Inverse Fourier--Bessel integration test.}
The regularized momentum-space companion uses the transform convention
\begin{equation}
f(x,k_T;Q)
=
\frac{1}{2\pi}\int_0^\infty db_T\, b_T\,
J_0(k_T b_T)\,
\widetilde f(x,b_T;Q),
\end{equation}
with inverse
\begin{equation}
\widetilde f(x,b_T;Q)
=
2\pi\int_0^\infty dk_T\, k_T\,
J_0(k_T b_T)\,
f(x,k_T;Q).
\end{equation}
Because the numerical grids are finite, the inverse transform is tested in the
active region rather than by demanding a pointwise relative error where the
function is numerically negligible.  The analytic test uses
\begin{equation}
\widetilde f_{\rm G}(b_T)=\exp(-a b_T^2),
\qquad a=0.45,
\end{equation}
whose forward transform is
\begin{equation}
f_{\rm G}(k_T)
=
\frac{1}{4\pi a}
\exp\!\left[-\frac{k_T^2}{4a}\right].
\end{equation}
With \(6001\) points in \(b_T\) up to \(30~{\rm GeV}^{-1}\) and \(8001\)
points in \(k_T\) up to \(80~{\rm GeV}\), the maximum absolute forward-transform
error for \(k_T\le12~{\rm GeV}\) is \(3.3\times10^{-7}\).  Transforming forward
and then back gives a 90th-percentile relative error of \(3.4\times10^{-3}\) for
\(b_T\le4~{\rm GeV}^{-1}\), and an active-region RMS relative error
\(6.8\times10^{-4}\).  The largest pointwise relative errors occur only near the
edge of the small tail, where relative errors are a poor measure of the
absolute transform accuracy.

The same forward--inverse test was repeated on the combined-ensemble median $b_T$-space
curves at \(Q=7.5~{\rm GeV}\), using the \(x\widetilde f\) curves for the six
light quark and antiquark flavors at \(x=0.1,0.2,0.3,\) and \(0.5\).  The curves
were extended with the same exp-\(b^2\) large-\(b_T\) continuation used in the
regularized \(k_T\)-space construction, transformed to \(k_T\) space, and then
transformed back to \(b_T\) space.  The comparison was restricted to active
points with \(b_T\le4~{\rm GeV}^{-1}\).  Across the 24 tested TMD curves,
the median curve-level 90th-percentile relative error is
\begin{equation}
3.9\times10^{-4},
\end{equation}
and the largest curve-level 90th-percentile relative error is
\begin{equation}
1.4\times10^{-3}.
\end{equation}
The largest pointwise active-region error over the tested TMD curves is
\(5.7\times10^{-3}\).  These values are well below the combined envelope
and are small compared with the systematic effects discussed elsewhere in this
section.

\paragraph*{Interpretation for the present extraction.}
The combined closure study supports three conclusions.  First, a prediction
inserted back as pseudo-data closes exactly at the data-point level, so the nominal observable convention is self-consistent.  Second, pseudo-data generated at the
specified uncertainty scale gives unit-normal pulls and correct Gaussian
coverage, including when data-set-wide normalization shifts are included.  Third,
the numerical Fourier--Bessel transform and its inverse are accurate at the
sub-percent level in the active $b_T$ region used for physics interpretation.
Together with the 50-member pseudo-data fit convergence test, these results
support the experimental component of the combined ensemble.  The comparison of independently initialized fits separately establishes the stability of the initialization
component.  Neither result assigns a calibrated confidence level to the
combined $q_{16}$--$q_{84}$ envelope.  They do not remove the need
for separate physics systematics from covariance modeling, collinear PDFs,
fiducial acceptance, perturbative profiles, or model-form variations.

\subsection{Covariance, PDF, and model-form variations}

A uniform full covariance representation is not available for all collider
inputs.  The nominal fit therefore uses the published tabulated uncertainties,
the stated point-to-point prescriptions, and normalization nuisances.  Future
analyses should compare this diagonal-plus-normalization treatment with
correlated-systematics implementations, especially for CDF Run II and the LHCb
fiducial spectrum.  In parallel, the large high-$x$ PDF effects observed in
selected fixed-target panels should be studied with multiple PDF sets and, where
practical, by PDF-through-refit rather than only PDF-through-reconstruction.  The
nominal result already includes separate tests of penalties on functional path
length, deviations in $\log F_{\rm NP}$, curvature, and the large-$b_T$ tail;
none reduced the initialization dependence while preserving the fit quality
sufficiently to replace the nominal objective.  Further model-form variations
include flavor-dependent $\FNP^q$, alternative architectures, scale/profile
variations, recomputing the reference curve after excluding each fit in turn,
and nuclear-target prescriptions for the fixed-target data.

\section{Discussion and limitations}
\label{sec:discussion}

The result supports three principal conclusions.  First, a compact, constrained
DNN can replace a low-dimensional nonperturbative ansatz while preserving the
perturbative transparency of $b_T$-space factorization.  The monotone
construction prevents the network from learning oscillatory compensation
patterns in $F_{\rm NP}$, and the FiLM conditioning captures smooth $x$
dependence while leaving the PDFs, hard coefficient, matching, evolution, and
Bessel kernel outside the trainable model.

Second, the $\NthreeLL$ label applies only to the explicitly separated resummed
$W$-term evolution with the stated strict-NLO hard/OPE matching.  The validation
tests examine the Born normalization, general-scale matching, profile behavior,
singular subtraction, and representative fixed-order tail conventions.  The
neural factor does not alter the perturbative accuracy assigned to the finite-$Y$
or larger-$q_T$ contributions.

The finite-$Y$ boundary comparison tests the transition from the resummed
region toward larger transverse momentum using 24 additional Tevatron points.
The nominal 329-point sample remains $W$-only, and the four higher-$q_T$ LHCb
points remain excluded.  The results with $F_{\rm NP}$ held fixed show that the
extracted nonperturbative factor is stable under the specified transition.

The subsequent direct Tevatron grid provides a direct perturbative $W+Y$
benchmark, while the coupled 329-point candidate tests a different question:
how the flexible nonperturbative extraction responds when the specified
finite-$Y$ inputs and stronger large-$b_T$ controls are introduced
simultaneously.  The
candidate narrows one large-$b_T$ region but broadens the data-sensitive core and
intermediate-$k_T$ domain, and it does not satisfy the same convergence
certificate as the nominal ensemble.  These studies therefore neither replace
the nominal extraction nor constitute a global $W+Y$ calculation with complete
covariance, perturbative-order, and matching uncertainties.

Third, the uncertainty construction separates the experimental-residual and
initialization components.  The experimental-residual central 68\% interval,
defined by $q_{16}$ and $q_{84}$, describes the 50-member residual ensemble
generated with the adopted effective-error model.  The initialization
envelope describes nonuniqueness within the specified initialization and
optimization family.  The combined $4{,}800$-member envelope propagates both
sources of spread but has no calibrated confidence-level interpretation.  The
near-unity 96-to-48 width ratios show stability under doubling the ensemble of independently initialized fits; they do not establish completeness against other model classes
or initialization families.

The principal limitations are equally important.  The 5\% point-to-point
component for E772 and E288-400 is a sensitivity prescription rather than a full
covariance model.  The Tevatron measurements use published tabulated
uncertainties as diagonal point-to-point errors together with normalization
nuisances where a complete covariance representation is unavailable.  The LHCb
input uses a DYTurbo-derived fiducial acceptance correction.  Collinear-PDF
uncertainty is estimated through reconstruction and selected $W(b_T)$
recalculations rather than a full PDF-through-refit ensemble.  The
nonperturbative factor is flavor independent.  The empirical reference is
constructed once from the full ensemble of independently initialized fits rather
than recomputed after excluding each fit in turn.  Scale and profile variations,
nuclear-model
variations, complete covariance information, fiducial-acceptance variations,
alternative architectures, flavor dependence, and alternative reference
constructions are not combined in the displayed envelope.  Higher-$q_T$ points
outside the factorization-validity range remain outside the nominal fit.  The
exploratory $W+Y$ candidate additionally changes the perturbative input and the
reference-distance prescription at the same time, so it cannot isolate a causal
finite-$Y$ shift without a matched $\lambda_{\rm ref}=1$ comparison.

An instructive methodological contrast emerges between the present
impact-parameter-space extraction and our previous direct-$k_T$ analysis
\cite{FernandoKeller2025}.  In the $k_T$ study, soft pointwise anchors, moment
constraints, curvature regularization, and large-$k_T$ tail controls were
introduced during model development to condition the inverse convolution, but
were removed sequentially
after closure and stability had been established.  The final profiles remained
smooth and well behaved under these ablations, with only the structural normalization imposed separately at
each tabulated $(x,Q)$ point and the positive, monotone representation retained.  In
the present $b_T$ analysis, by contrast, stable high-quality solutions required
retaining both the structural conditions
$F_{\rm NP}(x,0)=1$, $F_{\rm NP}(x,b_T)>0$, and
$\partial F_{\rm NP}/\partial b_T\leq0$, and a nonzero empirical
reference-distance term; a more appreciable initialization-dependent spread
also remained.  One possible contributor is that the momentum-space FiLM
network was conditioned explicitly on both $x$ and $Q$, so that the hard scale
provided an additional conditioning coordinate for the radial $k_T$
representation.  Here, $F_{\rm NP}(x,b_T)$ is conditioned directly on $x$,
whereas its scale dependence enters through the fixed perturbative kernel and
the subsequent Fourier--Bessel projection.  More generally, learning a
momentum-space profile that enters the measured $q_T$ convolution directly may
offer identifiability advantages for the inverse problem.  This comparison is
not controlled, however, because the two analyses differ in their data
coverage, perturbative content, forward operators, and optimization
objectives.  It therefore does not establish an intrinsic advantage of either
representation, but motivates matched studies in which these ingredients are
held fixed while only the $k_T$- versus $b_T$-space representation and the
conditioning variables are varied.

\section{Conclusion}
\label{sec:conclusion}

We have presented a physics-informed $b_T$-space extraction of unpolarized
TMDPDFs from a 329-point fixed-target and collider Drell--Yan data set.  The
perturbative $W$ term uses $\NthreeLL$ evolution with strict-NLO hard/OPE
matching, while a compact FiLM network learns only a positive-rate, monotone,
light-flavor-shared nonperturbative factor.  The fitting objective contains the
data residuals, normalization nuisances, and a direct-$F_{\rm NP}$ empirical
reference-distance term with unit weight over
$0.1\le b_T\le2.0~\GeV^{-1}$.  Under the effective uncertainty model, the
unpenalized data contribution is
$\chi^2_{\rm data}/N_{\rm acc}\simeq0.418$ for $N_{\rm acc}=329$
($\chi^2_{\rm data}\simeq137.46$); the normalization and reference terms are
kept separate.

The uncertainty analysis uses 96 converged, independently initialized fits and 50
experimental-residual fields.  Their combination contains $4{,}800$ members per
flavor but does not represent $4{,}800$ independent fits.  The combined result is
reported as an empirical $q_{16}$--$q_{84}$ envelope.  Experimental, initialization, combined, and
PDF uncertainties are displayed separately.

The resulting $b_T$-space TMDs are smooth and finite within the specified model.
Their light-flavor differences arise from the collinear PDFs, matching, and
evolution rather than from a flavor-dependent $F_{\rm NP}$.  The low-$b_T$
hook and mild enhancement are traced to the perturbative profile, OPE matching,
and evolution factors, not to oscillatory nonperturbative learning.  The
$k_T$-space distributions are regularized finite-$b_T$ Hankel transforms and
should not be interpreted as independent fits or fixed-order high-$k_T$
predictions.

A separate finite-$Y$ study extends the tested Tevatron kinematics to
$q_T/Q\simeq0.30$ using 24 additional points.  With $F_{\rm NP}$ held fixed,
early, central, and late transition profiles satisfy the convergence criterion
and change the nominal-sample predictions by at most approximately $1.2\%$.
Allowing $F_{\rm NP}$ to vary does not satisfy the convergence requirement and
is not used to define the finite-$Y$ robustness result.  Thus the boundary
comparison supports the stability of the extraction without changing the
nominal 329-point fit or its $W$-term accuracy statement.

An independent 122-bin Tevatron calculation also provides a finite, positive
direct $\NthreeLL+\mathrm{NNLO}$ $W+Y$ benchmark.  A separate candidate refit
using the specified non-LHCb finite-$Y$ inputs shows that extending the
reference-distance control to the full $b_T$ range relocates rather than removes
the initialization dependence.  Because this candidate changes several inputs
simultaneously and many optimizations terminate at the epoch ceiling, it is
retained as an identifiability diagnostic rather than promoted as a replacement
for the nominal $W$-only extraction.  The code and
numerical material are available at \url{https://github.com/uva-spin/b-space}.

\begin{acknowledgments}
This work used LHAPDF and the NNPDF4.0 PDF ensemble. The authors acknowledge Research Computing at the University of Virginia for providing computational resources and technical support that have contributed to the results reported in this publication. For additional information, see \url{https://rc.virginia.edu}. This work was supported by the U.S. Department of Energy under contract DE-FG02-96ER40950.
\end{acknowledgments}

\appendix

\section{Network optimization and numerical implementation}
\label{app:network}

\subsection{Architecture and numerical grids}

The network uses radial width 48, conditioning width 32, three FiLM residual
blocks, a positive Softplus head, and the integrated form in
Eq.~\eqref{eq:FNP}.  The cross-section calculation uses 160 points on a $b_T$
grid extending to $8~\GeV^{-1}$, and the tabulated $b_T$-space TMDs use 321
points over the same range.  The empirical-reference term uses the eight $x$
values in Eq.~\eqref{eq:reference-x-grid} and the fixed $b_T$-node set
$\mathcal B_{\rm ref}$ defined in Eq.~\eqref{eq:reference-b-domain}.  The regularized transform uses 6001 extended
$b_T$ points and 401 $k_T$ points from 0 to $4~\GeV$, as described in
Appendix~\ref{app:hankel}.

The numerical arrays underlying the figures are distributed with the
accompanying numerical material.  The fixed coordinates are $x=0.1$ and
$Q=7.5~\GeV$ in Fig.~\ref{fig:bspace}, $x=0.1$ and $Q=10~\GeV$ in
Fig.~\ref{fig:kspace-fixedx}, and $Q=7.5~\GeV$ in
Figs.~\ref{fig:kspace-surface-u} and \ref{fig:kspace-d}.  The exploratory
$W+Y$ surfaces in Figs.~\ref{fig:wy_candidate_u_surface} and
\ref{fig:wy_candidate_d_surface} use direct numerical results at $x=0.1$, 0.2,
0.3, and 0.5; the denser $x$ mesh is a display interpolation, and the
accompanying numerical tables retain the unclipped $q_{16}$, $q_{50}$, and
$q_{84}$ values.

\subsection{Optimization and convergence}

The independent optimizations begin from a common fitted initialization, after
which the trainable parameters are perturbed using independent random seeds.
Convergence is assessed from the block-to-block variation of $F_{\rm NP}$ on
$\mathcal X_{\rm ref}\times\mathcal B_{\rm ref}$.  A solution is retained only when the maximum relative drift
satisfies Eq.~\eqref{eq:convergence-criterion} for the specified number of
consecutive blocks after the minimum training exposure.  The ensemble contains
96 converged solutions; all 48 solutions added when the ensemble was enlarged
from 48 to 96 satisfy the same criterion.

The complete run configuration, including the optimizer, learning-rate schedule,
random seeds, perturbation amplitude, block definition, stopping tolerances,
separate objective contributions, and numerical plotting grids, is supplied
with the code and numerical material at
\url{https://github.com/uva-spin/b-space}.

\subsection{Experimental-residual combination}

The experimental component contains 50 experimental-residual fields obtained
with the same data selection, effective uncertainties, normalization treatment,
perturbative calculation, and factorization-validity prescription as the nominal
fit.  Each field is combined with each of the 96 converged, independently initialized fits, yielding $4{,}800$ members per flavor.  The construction contains 96
fitted networks and 50 residual fields.

\section{Experimental-residual ensemble checks}
\label{app:experimental-residuals}

For the 50 pseudo-data fits on the nominal 329-point sample used to construct
the experimental-residual fields, $\overline\chi^2$ denotes the mean squared
standardized residual
defined in Sec.~\ref{sec:validation}.
\begin{align}
\mathrm{median}(\overline\chi^{2})&=1.452069,\\
q_{95}(\overline\chi^{2})&=1.602821,\\
\max(\overline\chi^{2})&=1.728032,\\
q_{95}(|p_{\rm norm}|)&=2.970901,\\
\max(|p_{\rm norm}|)&=3.537025.
\end{align}
The largest normalization pull remains below the adopted threshold of 4.
The random-split center and width criteria are satisfied for the tabulated active-region
$b_T$-space curves.  These numbers characterize the 50-member experimental component only.

\section{Regularized Hankel transform}
\label{app:hankel}

The transformation first interpolates each $b_T$-space curve with a shape-preserving cubic interpolator.  Beyond the numerical grid endpoint $b_T^{\rm end}$, the default quadratic-exponential continuation is
\begin{equation}
\widetilde f(\bt)
=
\widetilde f(\bt^{\rm end})
\exp\left[
-a\left(\bt^2-(\bt^{\rm end})^2\right)
\right],
\end{equation}
where $a$ is fitted to the logarithm of the final segment in $b_T$ space.  A cosine taper multiplies the final 8\% of the extended transform grid.  The transform is evaluated with 6001 points in $b_T$ space and 401 points in $k_T$ space from 0 to $4~\GeV$.

The regularization comparison uses the median curves and the active-region criterion of Eq.~\eqref{eq:active-region}, with $z=\kt$.  Relative differences among the quadratic-exponential, exponential-in-$b_T$, and taper-only continuations are summarized by the median, 90th percentile, and maximum over this region.

\bibliographystyle{apsrev4-2}
\bibliography{bspace_PRD_references}

\end{document}